# A Unified Metric Architecture for AI Infrastructure: A Cross-Layer Taxonomy Integrating Economics, Performance, and Efficiency

Qi He[1]

Nov 20, 2025

## Abstract

The growth of large-scale AI systems is increasingly constrained by fundamental infrastructure limits. These constraints, including power availability, thermal and water constraints, interconnect scaling, and rapidly escalating lifecycle cost, interact across hyperscale clusters, yet the main metrics remain fragmented. Across hyperscale clusters, these constraints interact, yet the main metrics remain fragmented. Existing metrics, ranging from facility measures (PUE) and rack power density to network metrics (all-reduce latency) and financial metrics (TCO series), each capture only their own domain. This fragmentation obscures the structural relationships among energy, computation, and cost, preventing systemwide optimization and concealing how bottlenecks emerge, propagate, and jointly determine the efficiency frontier of AI infrastructure.

This paper develops an integrated framework that unifies these disparate metrics through a three-domain semantic classification and a six-layer architectural decomposition, producing a 6×3 taxonomy that maps how various constraints propagate across the AI infrastructure stack. The taxonomy is grounded in a systematic review and meta-analysis of all metrics with economic and financial relevance, identifying the most widely used measures, their research intensity, and their cross-domain interdependencies. Building on this evidence base, the Metric Propagation Graph (MPG) formalizes cross-layer dependencies, enabling systemwide interpretation, composite-metric construction, and multi-objective optimization of energy, carbon, and cost.

The framework offers a coherent foundation for benchmarking, cluster design, capacity planning, and lifecycle economic analysis by linking physical operations, computational efficiency, and cost outcomes within a unified analytic structure.

[1] Qi He is with Google LLC, Austin Office, USA. Email: lizziehe999@gmail.com.*The views expressed in this paper are solely those of the author and do not necessarily reflect the views of Google LLC or any other current or former employer or affiliated institution.*

# Introduction

As model scale grows, infrastructure, not only algorithms, has become a dominant factor influencing the efficiency, economics, and environmental footprint of AI systems. Training runs regularly consume tens of megawatt-hours, accelerator clusters operate at grid-relevant scales, and capital expenditure for state-of-the-art AI facilities now rivals that of utility-scale energy assets. Yet despite the speed and scale of this transformation, the AI infrastructure ecosystem lacks a transparent, coherent, and quantitatively grounded framework for measuring efficiency, reliability, and economic performance. Industry standards have not kept pace: Power Usage Effectiveness (PUE) remains the dominant facility benchmark even though it omits workload behavior; MLPerf captures workload performance but ignores energy and carbon; and total cost of ownership (TCO) models vary widely across vendors, often lacking methodological consistency. As a result, infrastructure decisions frequently rely on partial or incomparable metrics that obscure tradeoffs and hinder evidence-based planning.

This fragmentation arises because existing measurement traditions evolved independently. Facility engineers employ physical-efficiency ratios, xUE metrics (PUE, WUE, CUE, ERE), that quantify pre-IT thermodynamic losses but cannot represent algorithmic or economic behavior. Compute architects and ML systems engineers focus on FLOPs/W, utilization efficiency, communication overhead, and MLPerf convergence time, metrics that reveal software-hardware interaction yet lack any connection to energy procurement, facility overhead, or reliability risk. Economists and operators use TCO series, LCOE, and LCOH to evaluate financial viability, but these indicators do not reveal where inefficiencies originate or how they propagate across the system stack.

These metrics are not just disconnected, they are fundamentally incommensurable, defined in different dimensional systems (physical ratios, computational throughput, or monetary cost). Without a structural framework linking them, the industry faces: (1) incompatible benchmarks across vendors and regions, (2) inconsistent cost and carbon projections, (3) inability to compare alternative siting or cooling designs, (4) and no standardized way to evaluate hardware generations, cluster topologies, or runtime strategies. In a field where infrastructure spending is measured in billions of dollars and energy consumption is measured in terawatt-hours, the absence of a unified metric foundation has become a first-order scientific and economic problem.

This paper addresses this gap by introducing the first unified framework that connects physical, computational, and economic metrics of AI infrastructure into a single, interpretable system. My contributions are fourfold and concretely address long-standing blind spots in the literature, First, I formalize three meta-domains that capture the full semantic structure of AI-infrastructure metrics. Rather than grouping metrics by organization (e.g., “data center team” or “ML team”), I reveal that all established metrics fall into three scientifically grounded domains:

- physical facility efficiency (xUE family),
- compute and workload efficiency (FLOPs/W, tokens/J, MLPerf), and
- economic and reliability efficiency (TCO, LCOE, MTBF/MTTR).

This decomposition clarifies why cross-domain comparisons currently fail and provides a principled basis for integration.

Second, I introduce a six-layer architectural hierarchy representing the structural flow of energy, performance, and cost. This hierarchy, grid, facility, compute hardware, interconnect, ML runtime, and service economics, establishes the physical pathways through which inefficiencies propagate. It explains, for example, how grid carbon intensity affects training cost, how cooling topology shapes GPU reliability, and why network congestion increases TCO-per-token. This framework generalizes and formalizes the fragmented operational models used across hyperscalers.

Third, I develop the Metric Propagation Graph (MPG), a graph-theoretic representation of cross-layer dependencies. The MPG provides:

- interpretability (identifying bottlenecks and leverage points),
- metric synthesis (constructing carbon-normalized throughput, reliability-adjusted TCO, PUE-adjusted cost-per-token, etc.),
- optimization (quantifying tradeoffs among density, cooling, networking, carbon, and cost).

This tool transforms heterogeneous metrics into a single analyzable structure.

Fourth, I deliver a unified 6×3 taxonomy that operationalizes the entire metric landscape. Every metric, historical, contemporary, or emerging, maps precisely to one cell in this taxonomy. This enables transparent cross-site comparison, standardization of reporting, reproducible benchmarking, and consistent evaluation of future hardware, cooling, or siting strategies. In effect, the taxonomy provides the conceptual backbone for a unified AI-infrastructure standard.

Collectively, these contributions move beyond isolated metric definitions to offer a system-level, quantitative foundation for AI infrastructure, a foundation that is urgently needed as computational demand accelerates, energy constraints intensify, and capital investment grows. By bridging physical laws, algorithmic behavior, and financial dynamics, our framework provides a common language for engineers, economists, policymakers, and researchers who must jointly navigate the future of AI infrastructure.

The remainder of this paper is organized as follows. Section II reviews prior measurement traditions and their limitations. Section III formalizes the three meta-domains. Section IV presents the six-layer architecture. Section V introduces the Metric Propagation Graph. Section VI demonstrates systemwide applications. Section VII discusses implications. Section VIII concludes.

## Section II. Review of Existing AI Infrastructure Metrics

The landscape of AI infrastructure metrics reflects the convergence of several historically independent engineering and computational traditions, each shaped by its own objectives, measurement boundaries, and disciplinary assumptions. As modern AI workloads scale to trillion-parameter models and hyperscale clusters operating at tens or hundreds of megawatts, this fragmentation increasingly obstructs economic analysis, lifecycle planning, and system-level cost optimization. Most existing metrics were designed to assess isolated subsystems rather than integrated economic performance, making it difficult to link physical efficiency, workload behavior, reliability, and grid conditions to financial outcomes. This section reviews the major families of AI infrastructure metrics through the lens of economic relevance, focusing on their implications for capital expenditure (CapEx), operational expenditure (OpEx), lifecycle cost, risk exposure, productivity, and sustainability-linked financial impacts.

It is useful to orient the reader to the dominant metric families referenced in both industry and academic work. Each family originates in a distinct disciplinary community and evaluates a different operational boundary of AI systems, from facility-level thermodynamics to ML workload execution to financial cost accounting. Table 1 summarizes these families, their origins, and representative indicators. The focus of this review is not exhaustiveness, but on the subset of metrics with direct or indirect implications for the economics of AI infrastructure.

Table 1. Economically-Relevant Families of AI Infrastructure Metrics and Their Technical Origins

| Metric Family | Originating Discipline | Economic / Financial Relevance | Measurement Boundary | Representative Metrics (Examples) |
|---|---|---|---|---|
| **Facility Efficiency & Resource Overhead Metrics** | Data center engineering; ISO/IEC 30134; ASHRAE Datacom | Determines facility OpEx, cooling cost, energy procurement exposure, carbon-embedded cost | Facility physical infra (power, cooling, water, heat rejection) | PUE, WUE, CUE, ERE; Cooling Power Ratio; Thermal Design Margin |
| **Compute Efficiency & Device-Level Performance Metrics** | HPC benchmarking; GPU/TPU architecture; accelerator design | Directly shapes compute-per-dollar, job duration, amortization rate of CapEx, marginal cost of compute | Chip, GPU/TPU, node-level compute hardware | FLOPs/W, joules per token, tokens/J, accelerator utilization, thermal throttling indicators |
| **ML System / Workload Execution Metrics** | Distributed training systems; MLPerf; model-serving research | Determines productivity cost (developer-time), job completion time (affects OpEx), scheduling efficiency, cost/training-run | Job-level, pipeline-level, distributed runtime | MLPerf throughput; time-to-convergence; inference latency; inference/request; idle energy overhead |
| **Economic & Lifecycle Cost Metrics** | Cloud economics; IT finance; energy economics | Captures total lifecycle cost (CapEx/OpEx), depreciation, cost of energy, carbon-adjusted cost, and long-run economic sustainability | Organizational, cluster, region, and workload levels | TCO, LCOE, CapEx-to-OpEx allocation, depreciation curve, marginal cost of compute unit |
| **Reliability, Service Continuity, and Risk Metrics** | SRE engineering; reliability theory; operations research | Quantifies risk cost, SLA penalty exposure, redundancy overhead, downtime loss, risk-adjusted TCO | Cluster, fleet, service, and component levels | SLA/SLO compliance; MTBF, MTTR, restart cost, redundancy cost, expected downtime loss |
| **Grid Interaction & Sustainability-Embedded Economics Metrics** | Power systems engineering; carbon accounting; energy markets | Determines carbon-adjusted TCO, regional cost arbitrate, demand-response revenue, price volatility risk | Regional grid, power market interface | Marginal Emissions Factor (MEF); LMP/RTM electricity price; carbon price exposure; grid-interaction cost |

*Notes:* This table provides a high-level orientation to the metric families discussed in this section. Detailed critique and structural limitations of these families follow in the paragraphs below. Details of all the metrics are shown in Appendix A, Appendix C (PUE, WUE, CUE, ERE), and Appendix D (TCO series).

Independent of facility engineering, high-performance computing (HPC) developed its own metrics centered on device-level efficiency. Canonical measures, such as FLOPs per watt, joules per inference, tokens per joule, and energy per training step, characterize the energy and performance behavior of CPUs, GPUs, and AI accelerators. These indicators reflect hardware fundamentals including microarchitecture, memory hierarchy, interconnect performance, and thermal envelopes. Although they are not financial metrics in isolation, they determine effective compute-per-dollar, performance-per-watt, and ultimately the marginal cost of executing AI workloads. Inefficient compute translates into higher OpEx, accelerated hardware depreciation, and underutilized CapEx. The economic relevance of these metrics therefore arises through their influence on hardware TCO, utilization efficiency, and the cost of delivering a unit of effective computation.

A third and increasingly influential tradition arises from machine learning systems engineering, where workload-level metrics describe the combined behavior of software stacks, compiler optimizations, model architectures, and distributed execution strategies. The MLPerf benchmark suite, developed by a consortium including Google, NVIDIA, Meta, Stanford, and Harvard, has become the de facto standard for evaluating AI training and inference performance. MLPerf Training reports time-to-convergence under fixed accuracy targets, a metric with immediate economic implications for developer productivity, experimentation velocity, and time-to-insight. MLPerf Inference reports latency and throughput across a variety of server and edge scenarios; these measurements directly shape serving cost per request, autoscaling requirements, and cloud revenue efficiency. While MLPerf is largely orthogonal to physical efficiency or carbon intensity, it plays a central role in linking workload behavior to operational cost and service economics.

Economics introduces a different measurement tradition concerning explicitly with cost. Total Cost of Ownership (TCO) frameworks quantify the full lifecycle cost of deploying and operating infrastructure, including capital expenditure, operational expenditure, energy procurement, staffing, interconnect costs, hardware refresh cycles, and depreciation. They are widely used to evaluate server and facility investments in both industry practice and the academic literature on data-center cost modeling.[1,2] Parallel work in energy economics has produced levelized cost metrics such as the levelized cost of electricity (LCOE), which express discounted lifecycle energy cost per unit of delivered power and are increasingly applied to data-center energy solutions and siting decisions.[3]

In modern AI data centers, however, TCO is increasingly driven by a compact set of hardware and facility design primitives. First, accelerator cost and GPU density (GPUs per server chassis) define the base compute CapEx and allocating the accelerator capacity and its associated infrastructure CapEx (motherboard, chassis, and power domain) over its useful life; server acquisition cost and power consumption are repeatedly identified as dominant contributors to overall TCO.[1] Second, planned or operational power draw per server and per rack governs power-delivery and cooling provisioning, and thus the allocation of facility CapEx and energy-related OpEx to each compute unit, higher server and rack power densities tighten cooling constraints and increase the marginal cost of additional load.[2,4] Third, networking topology and uplink capacity determine the quantity and class of switches, cabling, and optical links required per accelerator, so that network design becomes a first-order determinant of both capital and energy cost in large cloud data centers.[4,5] Fourth, server density (servers per rack) interacts with power draw and spatial layout to set rack-level power density, white-space utilization, and the effective amortization of floor space and distribution infrastructure.[1,2] Fifth, memory, storage, and other major components (e.g., DRAM, SSDs, NICs) account for a substantial share of system power and bill-of-materials; recent work shows that memory systems alone can consume on the order of 25-40% of total data-center power, making capacity and bandwidth choices inseparable from cooling and TCO considerations.[6,7] Finally, the economic useful life of servers, accelerators, and facility infrastructure fixes the amortization horizon and refresh cadence, shaping the annualized cost profile over which these capital expenses are recovered.[2]

These primitives give TCO models their direct financial interpretability, but they also reveal an important limitation: efficiency, reliability, and grid conditions usually enter as exogenous technical inputs rather than endogenous variables that are jointly optimized with hardware density, networking design, and facility architecture. In effect, contemporary TCO practice maps heterogeneous engineering quantities into financial terms without resolving the fragmentation of the underlying technical metrics (see Appendix Table D), leaving a gap between subsystem-level measurements and holistic economic evaluation.

In contrast, reliability engineering focuses on operational continuity and risk. The Google Site Reliability Engineering (SRE) framework popularized indicators such as Service Level Indicators (SLIs), Service Level Objectives (SLOs), Service Level Agreements (SLAs), Mean Time Between Failures (MTBF), Time to Converge (TTC), and Mean Time to Repair (MTTR). Hyperscalers devote considerable attention to fleet-wide component failure rates, GPUs, CPUs, DIMMs, SSDs, NICs, and to thermal- and workload-induced failure patterns. These metrics have substantial economic implications: downtime incurs SLA penalties, degraded throughput increases serving cost, and high failure rates necessitate overprovisioning or redundant CapEx. Thus reliability metrics connect physical stressors and operational behavior to risk-adjusted cost, forming a bridge between engineering factors and financial exposure.

Finally, grid-coupled metrics emerge from power systems engineering and environmental accounting rather than computing. As AI clusters increasingly draw industrial-scale power, metrics such as marginal emissions factors, temporal carbon intensity, renewable penetration ratios, and demand-response participation rates have become essential for evaluating sustainability and cost exposure. Time-varying electricity prices (e.g., locational marginal prices) directly affect OpEx, while carbon intensity directly affects carbon-pricing risk and compliance cost. Although these metrics are external to the computing stack, they shape the economic operating environment of AI infrastructure and increasingly influence workload placement, scheduling, and site selection decisions.

Despite their internal coherence, these measurement traditions share no common semantics, units, or operational boundaries. A Power Usage Effectiveness (PUE) value cannot be converted into FLOPs per watt; an MLPerf score provides no insight into carbon cost; a TCO model cannot diagnose workload inefficiency; and server failure rates cannot be used to infer thermal overhead or long-term lifecycle cost. These incompatibilities are not merely conceptual but organizational. Facility teams optimize thermal and mechanical efficiency; hardware teams maximize accelerator performance and reliability; ML systems teams optimize throughput and convergence; economics teams optimize TCO and lifecycle investment; and sustainability teams monitor carbon intensity and regulatory exposure. Each group maintains its own key performance indicators (KPIs) and dashboards, which complicates system-wide economic reasoning.

The cumulative effect is a metrics landscape that is fundamentally unsuited to holistic economic evaluation. AI infrastructure metrics evolved within isolated disciplinary silos, not within a unifying economic framework. This fragmentation limits the ability of analysts to evaluate trade-offs, compare system configurations, optimize cost-performance ratios, or understand the economic consequences of design decisions across layers. These limitations motivate the framework introduced in the next sections, which formalizes three meta-level domains and decomposes AI infrastructure into six architectural layers. Together, they define a unified 3×6 taxonomy that integrates physical efficiency, workload performance, reliability, and grid behavior into a coherent structure for economic evaluation. To illustrate these divergences more concretely, Table 2 summarizes representative metrics from each family, emphasizing the dimensions most relevant to economic analysis, units, temporal scope, spatial scope, and semantic orientation.

Table 2. Representative Metrics in Major AI Infrastructure Metric Families with Various Characteristics

| Metric Family | Representative Metric | Units | Temporal Scope | Spatial Scope | Economic / Semantic Focus |
|---|---|---|---|---|---|
| **Grid & Sustainability** | Grid carbon intensity (CI); MEF | $gCO_2/kWh$ | hourly → sub-hourly | regional grid / power market | Carbon-adjusted cost of compute, electricity price, DR participation value |
| | | | | | |
| **Facility Efficiency** | PUE (Power Usage Effectiveness) | ratio (kWh/kWh) | hourly → monthly | datacenter / facility | Cooling OpEx, energy overhead cost, carbon-embedded cost; thermodynamic inefficiency pathways |
| **Compute Efficiency** | FLOPs/W or Tokens/J | FLOPs per watt; joules per token | microsecond → second | chip / GPU / TPU / rack | Compute-per-dollar, thermal throttling impacts on OpEx, accelerator amortization efficiency |
| **Workload Efficiency** | Joules per inference; Time-to-convergence | J/op; time to accuracy | millisecond → job / batch | job / pipeline / distributed cluster | Productivity cost, pipeline congestion, cost-per-inference, scheduling elasticity |
| **Economic Metrics** | TCO; LCOE; marginal cost of compute | USD/year; USD/op | annual → multi-year | cluster / region / org | Lifecycle finance (CapEx→OpEx), depreciation, price exposure, long-run cost sustainability |
| **Reliability & Operations** | SLA availability; MTBF/MTTR | % availability; hours | event-driven; daily | fleet / service / component | Downtime loss, SLA penalty risk, redundancy overhead, risk-adjusted TCO |

*Notes:* This table reports representative, not exhaustive, metrics within each family. Individual metrics may differ substantially in units, temporal granularity, and measurement boundaries. The table is intended to illustrate heterogeneity across families rather than internal completeness.

Despite substantial progress in developing individual metric series, the existing literature remains fragmented. Prior studies typically examine metrics within a single layer or domain and do not provide a unified analytical structure that spans facility, compute, workload, reliability, and economic dimensions. Moreover, existing frameworks do not offer a formal mechanism for composing heterogeneous metrics into coherent, economically interpretable quantities or for analyzing how variations at one layer propagate through the rest of the infrastructure. These gaps motivate the integrated three-domain decomposition (Section III), the six-layer architectural formulation (Section IV), and the unified 6×3 taxonomy and Metric Propagation Graph developed in Sections V and VI.

# SECTION III. Three Meta-Level Metric Domains

Modern AI infrastructure exhibits interactions among physical systems, computational workloads, and economic mechanisms that operate across incompatible units, boundaries, and modeling assumptions. The metric families historically used to evaluate these systems, facility metrics (e.g., PUE), compute metrics (e.g., FLOPs/W), workload metrics (e.g., MLPerf latency), and economic metrics (e.g., TCO), reflect independent scientific lineages. As shown in Section II, this fragmentation prevents integrated evaluation, joint optimization, and cost-performance tradeoff analysis. To enable system-level reasoning, I introduce three mutually exclusive, conceptually coherent, and operationally meaningful meta-metric domains: (1) Physical Facility Efficiency; (2) Compute & Workload Efficiency; (3) Economic & Reliability Efficiency

These domains define the semantic axes of the AI-infrastructure metric space and form the horizontal dimension of the unified 6×3 taxonomy (Section V). The domains are orthogonal (distinct physical, computational, and economic processes), complete (every known metric maps to exactly one domain), and minimal (no fourth domain is needed without redundancy).

## III.1. Domain 1. Physical Facility Performance and Efficiency (the xUE Family)

**Origins in Data Center Engineering, Thermodynamics, and ISO/IEC Standards**

This domain originates in data center engineering, thermodynamics, and ISO/IEC 30134 standards, which formalize PUE, WUE, CUE, and ERE [8]. Together with ASHRAE thermal guidelines [9], these standards define airflow efficiency, economization, COP, and other facility-level mechanisms that govern the conversion of utility power into usable IT load. These metrics are widely reported across hyperscaler disclosures (Google [10-12], Meta [13], AWS/Microsoft) and quantified across thousands of sites in Uptime Institute surveys.

The xUE family captures pre-IT physical loss mechanisms in units fundamentally incompatible with computational or economic metrics: kWh/kWh, L/kWh, $gCO_2$/kWh, °C/W. Although physical, these metrics directly influence OpEx (energy, water), facility CapEx (cooling infrastructure), and carbon-cost exposure.

### III.2. Domain 2. Compute & Workload Efficiency

**Origins in HPC, Accelerator Design, ML System Benchmarking, and Model Serving**

The second meta-domain emerges from HPC, accelerator design, and distributed ML systems. Green500 formalized FLOPs/W as the canonical compute-efficiency metric [14], and modern accelerator platforms (e.g., NVIDIA DGX, TPU Pods) extend this to tokens/Joule, joules per training step, tensor-parallel scaling efficiency, and communication overhead [15-16].

MLPerf Training and Inference benchmarks [17-18] introduce standardized workload-level metrics, time-to-convergence, latency, and throughput, which capture full-stack execution behavior not inferable from hardware-only measures.

This domain measures useful computational work in FLOPs, Joules/op, batches/s, sequences/s, units distinct from thermodynamic ratios or monetary cost. Compute inefficiency translates directly into higher cost-per-token, serving cost-per-request, and lower amortization efficiency of hardware CapEx.

### III.3. Domain 3. Lifecycle Economics & Reliability Risk

**Origins in Lifecycle Economics, TCO Models, SRE Framework, and Risk/Operations Research**

The third meta-domain arises from lifecycle economics, cloud TCO modeling, and SRE reliability engineering. TCO decomposes CapEx, OpEx, depreciation, energy, interconnect, and refresh cycles [19-21], while energy economics contributes LCOE and emerging LCOAI models [22-23].

Reliability metrics, MTBF, MTTR, SLOs, error budgets, stem from Google's SRE framework [24] and quantify failure risk, redundancy requirements, downtime loss, and SLA penalties. These metrics operate in USD, probability distributions, and depreciation rates, representing lifecycle and risk semantics rather than physical or computational behavior.

**Structural Rationale for the Three Meta-Metric Domains**

The three-domain decomposition is justified by three structural properties of AI infrastructure metrics. First, the domains correspond to fundamentally distinct scientific laws: physical metrics obey thermodynamic and mechanical constraints; compute metrics follow algorithmic and architectural scaling principles; and economic and reliability metrics arise from financial and stochastic models. No domain can be reduced to another without introducing an external modeling interface. Second, the domains operate in incompatible dimensional systems, physical ratios (e.g., kWh/kWh, L/kWh, $gCO_2$/kWh), computational units (e.g., FLOPs, Joules/op, latency), and monetary or probabilistic quantities (e.g., USD/year, LCOE, failure rates), which are mathematically incommensurable and cannot be jointly interpreted without semantic distortion. Third, the decomposition is both minimal and complete: every known AI-infrastructure metric maps uniquely to exactly one of the three domains and introducing a fourth domain would produce redundancy rather than analytical clarity. A full formal justification is presented in Appendix E.

**Empirical and Industry Support for the Three-Domain Decomposition**

Multiple independent literatures converge on this three-domain structure. Although hyperscalers often organize engineering teams into “Data Center,” “Server/Compute,” and “Network,” these organizational boundaries do not align with the underlying physical-computational-economic decomposition. The three-domain approach captures the actual flow of energy, workload execution, reliability dynamics, and cost evolution more accurately than organizational charts.

While the three meta-metric domains establish the semantic structure of the AI infrastructure metric space, they do not specify where in the system stack each metric operates. To connect domain-level semantics with system-level structure, a complementary architectural decomposition is required. Section IV introduces a six-layer framework that captures the vertical propagation of energy, performance, reliability, and cost through the AI infrastructure stack. When combined with the three meta-domains, this yields the unified 6×3 taxonomy used throughout the remainder of this paper.

# Section IV. The Six-Layer Architecture of AI Infrastructure

Modern AI systems operate at a scale where physical infrastructure, computational subsystems, and economic considerations interact in tightly coupled and often nonlinear ways. While Section III established the semantic structure of metrics through three meta-domains, physical facility efficiency, compute and workload efficiency, and economic and reliability efficiency, these domains alone do not specify the structural locations within the system where metrics arise. Because cost, performance, and sustainability evolve through cascaded interactions across different technical subsystems, a structural decomposition of the AI infrastructure stack is required. This section introduces a unified six-layer architecture that integrates energy systems, facility engineering, computing hardware, distributed cluster design, workload execution, and economic reliability mechanisms into a cohesive framework. The architecture extends, generalizes, and refines the five-layer operational models implicitly used by hyperscalers such as Google, Meta, NVIDIA, Microsoft, and AWS, while adding two layers, Grid & Sustainability, and Economic/Reliability, essential for cost modeling and lifecycle financial evaluation. A high-level representation of this architecture is shown in Figure 4.

## IV.1. The Six Layers

**(1) Grid & Sustainability Layer (Layer 1)**

At the foundation of AI infrastructure lies the electric power grid that supplies energy to data centers and AI compute clusters. This layer encompasses the physical and economic characteristics of electricity generation, transmission, carbon intensity, and grid stability. Key observables include marginal emissions factors, locational marginal pricing (LMP), renewable penetration, temporal carbon intensity, and grid congestion patterns. These variables originate from power-systems engineering rather than computing [26-29], yet they exert first-order impacts on AI operational cost and environmental performance.

Grid conditions determine the baseline marginal energy cost, which flows directly into compute energy expenditure and ultimately into lifecycle cost models such as TCO. In addition, sustainability and resilience levers can extend beyond day-to-day operations: for instance, modular landfill remediation can be co-modeled with grid-level metrics to improve AI corridor resilience while reducing methane-related externalities, yielding measurable Layer-1 impacts on emissions and reliability [57]. Marginal emissions factors determine the carbon intensity of compute, influencing carbon pricing exposure, sustainability targets, and the effective cost-per-inference under carbon-regulated environments. Time-of-use pricing structures propagate into workload scheduling decisions, for example, shifting batch inference to periods of low grid carbon intensity or low marginal price. Although hyperscaler operational teams rarely classify the grid as part of the infrastructure “stack,” the grid layer is indispensable for modern AI economics because it defines the initial conditions under which all downstream energy transformations occur.

**(2) Facility Layer (Layer 2)**

The Facility Layer encompasses all pre-IT mechanical, electrical, and thermal subsystems, cooling plants, air-handling units, power distribution paths, battery and UPS systems, heat rejection mechanisms, water consumption, and environmental control systems. The ISO/IEC 30134 family of international standards formalizes the principal facility-level efficiency ratios, including PUE, WUE, CUE, and ERE [8], while ASHRAE’s Datacom thermal guidelines establish temperature ranges and humidity constraints that shape overall thermal operating conditions [9]. These metrics reflect the ability of the facility to deliver stable electrical and thermal environments that enable reliable operation of compute hardware.

From an economic perspective, the facility layer has a dual influence. First, physical efficiency directly determines operational expenditure: a PUE of 1.2 implies that for each watt consumed by IT devices, 0.2 watts are lost to cooling and electrical overhead, magnifying energy cost by 20%. Second, thermal management affects the longevity and failure rates of hardware. Elevated temperatures increase the probability of GPU throttling, memory bit errors, and power-supply failures, ultimately driving up MTBF-adjusted replacement cost and influencing long-term TCO. The facility layer thus forms a critical conversion boundary between grid-supplied energy and usable IT power.

**(3) Compute Hardware Layer (Layer 3)**

The Compute Hardware Layer contains the physical accelerators and compute units, GPUs, TPUs, CPUs, ASICs, NPUs, memory subsystems, and power delivery mechanisms, that execute AI workloads. Metrics in this layer originate from HPC and accelerator research, including FLOPs/W, joules per operation, memory bandwidth efficiency, tensor core utilization, and tokens per joule [15-16]. These metrics capture how effectively compute hardware converts electrical energy into useful mathematical computation.

In economic terms, the compute layer is the core determinant of marginal cost-per-token for large-scale training and inference. Hardware utilization and energy-per-operation directly influence both immediate OpEx and amortized CapEx recovery. Underutilized accelerators prolong training time and reduce return on invested compute capital, while inefficient memory or interconnect behavior elevates the cost of serving inference. The compute layer also interfaces deeply with reliability economics: GPU failure rates, HBM degradation, and NIC failure behaviors influence replacement cycles and redundancy requirements, all of which enter lifecycle cost models.

**(4) Cluster Networking & Interconnect Layer (Layer 4)**

As modern AI workloads require distributed execution across hundreds or thousands of accelerators, the networking and interconnect layer plays a decisive role in determining overall system productivity. This layer includes high-performance fabrics such as InfiniBand, NVLink, Ethernet with RDMA, optical interconnects, topologies (fat-tree, dragonfly, torus), and congestion control protocols. Metrics in this layer reflect communication bandwidth, latency, collective operations overhead, network energy efficiency, and cluster-level communication balance.

Economically, the interconnect layer establishes the scaling efficiency of distributed training and inference. A system with high FLOPs/W at the hardware layer may nonetheless exhibit high cost-per-training-step if network congestion or topology imbalance forces synchronization delays. Each additional millisecond of all-reduce latency translates into longer job completion times, consuming more GPU-hours and increasing both OpEx and CapEx amortization. Thus, the interconnect layer is the primary mediator between hardware potential and workload-level realized performance.

**(5) ML Runtime, Workload Execution, and Reliability and Resilience Layer (Layer 5)**

Above the hardware and interconnect layers lies the ML runtime layer, which includes compilers, distributed training frameworks, runtime schedulers, and software execution engines. Key influences in this layer include data parallelism, tensor parallelism, pipeline parallelism, activation checkpointing, memory optimization passes, kernel fusion, and runtime load balancing. MLPerf Training and Inference benchmarks reflect this layer's behavior by measuring time-to-convergence, latency, throughput, and accuracy under standardized workloads [17-18]. Domain-specific engineering workloads, for example, GAN- and self-attention-based 3D evaluation of tunnel water leakage [51], also map naturally to this layer, where runtime efficiency and model quality are jointly measured.

This layer introduces a software-driven multiplier effect on economic outcomes. Inefficiencies in parallelization strategies or compiler optimizations can degrade utilization, extend training timelines, and increase inference serving cost-per-request. Conversely, improvements in runtime scheduling or model-parallel design can deliver cost savings far exceeding those achievable solely through hardware upgrades. As such, the ML runtime layer is the locus where algorithmic efficiency translates into economic efficiency.

**(6) Service, Operation, and Economic Layer (Layer 6)**

The topmost layer integrates service-level delivery, operational reliability, and full lifecycle economics. Economic models such as TCO, LCOE [19-23], depreciation schedules, discounting models, and risk-adjusted cost estimates quantify long-term financial exposure. In parallel, Google's SRE framework formalizes reliability metrics such as SLI/SLO/SLA compliance, error budgets,

MTBF, and MTTR [24], which influence service continuity and availability guarantees.

This layer is where cost, reliability, and performance converge, and where the economic consequences of all lower-layer behaviors are synthesized. For multi-campus fleets, this economic synthesis also depends on how site-level costs are aggregated and normalized for fair comparison and allocation; for example, a location-robust, cost-preserving blended pricing method can preserve total costs while avoiding rank-order artifacts across heterogeneous data centers [56]. Failures originating from lower layers propagate into SLA penalties, redundancy overhead, and downtime costs. Energy inefficiency from the grid or facility layers manifests as increased cost-per-inference or increased carbon pricing liability. Similarly, workload inefficiencies observed at the runtime layer translate into higher compute demand and extended resource usage. The Service & Economic Layer therefore functions as the integrative boundary where the economic consequences of all lower-layer behaviors are synthesized.

## IV.2. Cross-Layer Interactions and Economic Propagation Pathways

The six layers do not operate as isolated modules; instead, they form a tightly interconnected chain where physical, computational, and economic effects propagate upward and downward. Thermal inefficiencies at the facility layer can induce GPU throttling, reducing throughput at the compute layer, amplifying synchronization delays at the interconnect layer, and ultimately increasing cost at the service layer. Similarly, workload imbalance at the runtime layer pushes unnecessary data movement across the interconnect fabric, elevating energy draw and increasing carbon intensity at the grid layer. Reliability degradations propagate across layers as well: an elevated failure rate in the compute layer necessitates redundancy, which increases energy use and accelerates depreciation in the layer.

These interactions form what can be described as an energy-performance-reliability-cost chain, where disturbances at any layer produce quantifiable economic outcomes. The layered architecture therefore provides a structured view of how metrics at different layers influence one another, establishing a foundation for the unified taxonomy in Section V.

Although hyperscaler organizations often conceptualize their infrastructure using a five-layer structure that includes facility engineering, server hardware, networking, runtime systems, and service delivery, this operational model omits two critical layers essential for economic and lifecycle modeling: the grid layer and the explicit economic/reliability layer. The six-layer architecture presented here generalizes the hyperscaler view by explicitly incorporating upstream energy systems and downstream cost modeling. This expanded structure remains fully compatible with industrial mental models while enabling deeper technical analysis, particularly for economic and sustainability analytics.

With the six layers providing the structural axis of AI infrastructure and the three meta-domains defining the semantic axis of metrics, I now combine these two dimensions into a unified 6×3 taxonomy. This taxonomy situates each metric at the intersection of its system location and its semantic domain, enabling cross-layer analysis, economic reasoning, and systematic comparison across diverse AI infrastructures. Section V presents this unified metrics framework in detail.

# Section V. The Unified 6×3 Metrics Taxonomy

Modern AI infrastructure spans physical facilities, compute hardware, distributed interconnects, workload execution frameworks, and economic service layers. As demonstrated in Sections III and IV, existing metric families originate from distinct engineering traditions and occupy different structural layers within the system stack. However, without a unifying analytical framework, these metrics remain fragmented, preventing cross-layer reasoning, cost decomposition, or systematic comparison of alternative AI architectures. To address this gap, this section introduces a Unified 6×3 Metrics Taxonomy that positions every metric at the intersection of (i) one of six system layers and (ii) one of three meta-domains. This structure provides a comprehensive and interpretable mapping of the metrics universe for AI infrastructure.

## V.1. A Bidimensional Coordinate System for AI Infrastructure Metrics

The unified taxonomy is built as a two-dimensional coordinate system.
The vertical axis consists of the six structural layers established in Section IV. The horizontal axis consists of the three meta-domains introduced in Section III. Together these axes form an 18-cell (6×3) metrics universe, where each cell represents a unique type of system behavior. Every metric, regardless of whether it originates from HPC, MLPerf, ISO/IEC facility standards, power systems engineering, or cloud economics, can be unambiguously assigned to exactly one cell. To ground this structure, Table 3 provides the complete 6×3 classification of economically relevant metrics aligned with the six-layer architecture.

Table 3. Economically-Relevant Metrics Across Three Meta-Domains and Six Infrastructure Layers

| Layers | Domain 1: Physical Performance and Efficiency | Domain 2: Compute & Workload Efficiency | Domain 3: Lifecycle Economics & Reliability Risk |
|---|---|---|---|
| **1. Grid & Sustainability Layer** | Marginal Emissions Factor (MEF);<br>Time-varying grid carbon intensity;<br>Renewable availability | Grid-constrained Workload Shiftability;<br>Carbon-aware Scheduling Limits | Time-varying Electricity Price (LMP/RTM); Carbon Pricing;<br>Location-based Energy Arbitrage;<br>Grid Interaction Cost (DR programs) |
| **2. Facility Layer** | XUE Family (PUE, WUE, CUE, ERE) ;<br>OpEx, carbon price exposure;<br>Cooling Power Ratio;<br>Energy Reuse Ratio;<br>Thermal Design Margin (cooling CapEx) | Inefficient facility-level scheduling impacting job latency;<br>Facility congestion affecting throughput efficiency;<br>Temperature-dependent Throttling Constraints | Energy Opex Share (% of OpEx);<br>Cooling Opex;<br>Carbon Cost from CUE × Carbon Price;<br>Demand-Response Revenue Potential |
| **3. Compute Hardware Layer** | Device Power Draw (GPU/TPU);<br>Temperature → Performance Coupling;<br>Thermal Headroom | FLOPs/W;<br>Joules per Operation;<br>Accelerator Utilization;<br>Tokens/Joule | Hardware TCO Contribution;<br>Depreciation Curve (Lifetime);<br>Cost per Accelerator-Hour;<br>Failure-driven Replacement Cost |
| **4. Networking & Interconnect** | Network Power Overhead (Switch + NIC Cooling);<br>Rack/Pod Thermal Load | Pipeline-throughput → amortization cost;<br>Bandwidth/Watt;<br>Collective Operations Latency;<br>Congestion-induced Inefficiency | Interconnect CapEx;<br>Network Scaling Cost;<br>Topology-driven Throughput Loss |
| **5. ML Execution & Reliability Layer** | Thermal-induced failure rate;<br>Power-quality-induced failure;<br>Compiler-induced Energy Overhead;<br>Idle Energy from Inefficient Scheduling | MLPerf Metrics (Time-to-Convergence, Throughput, Latency);<br>Tokens-per-second Efficiency;<br>Model/Parallelism Efficiency;<br>Checkpoint Energy Performance degraded QoS | Cost per Training Step;<br>Cost per Token / per Inference;<br>Serving Cost Elasticity;<br>Autoscaling Cost Models |
| **6. Service, Operations & Economic Layer** | Redundancy Energy Overhead;<br>Service-layer Thermal Penalties | SLI/SLO Reliability Metrics;<br>MTBF/MTTR-driven Availability;<br>Latency QoS Compliance Overhead | TCO, LCOE, LCOC;<br>SLA Penalty Cost;<br>Downtime Loss;<br>Risk-adjusted Lifecycle Cost;<br>Expected Loss from MTBF/MTTR |

*Notes:* This matrix includes only metrics with direct or indirect economic consequences. Metrics are retained if they influence CapEx, OpEx, lifecycle cost, marginal compute

cost, failure cost, or sustainability-linked financial exposure.

## V.2. A Conceptual Overview

(1) Physical-domain metrics across Layers 1-6

These metrics capture fundamental physical flows, energy, water, heat, carbon, and power quality. Similar multi-scale carbon-sink accounting frameworks have been developed in land-use and agroforestry systems, where explicit and implicit carbon flows under cropland transformation are jointly evaluated [50]. Examples include grid carbon intensity (Layer 1), PUE/WUE/CUE (Layer 2), thermal-power coupling in GPUs (Layer 3), network cooling overhead (Layer 4), and service-layer redundancy energy (Layer 6). They quantify thermodynamic properties that impose constraints and costs on the compute system but do not represent computational work.

(2) Compute and workload-domain metrics across Layers 1-6

These metrics characterize the productive computational behavior of the system, including FLOPs/W, joules/op, MLPerf throughput, congestion-induced performance penalties, parallelism efficiency, checkpoint overhead, and latency. They quantify useful computational output and map directly to job completion time, trainer productivity, and inference efficiency, often mediated by memory- and coordination-related bottlenecks in modern serving stacks (e.g., KV-cache effects in LLM inference [62] and scheduling/coordination overheads in mixture-of-experts inference [63]).

At the workload level, “performance” is still task-defined: evaluations may include lightweight interpretability checks [52], data-sample prioritization for training efficiency [66] and long-tail multi-label learning performance [67], and domain KPIs such as tracking quality [54], [55], structured perception/mapping accuracy in autonomous driving [58] (including efficient traffic sign detection [69]) and automated waste monitoring in smart cities [68], robustness to manipulations in forgery detection [59], and cross-domain retrieval ranking quality and robustness [60], [61], including text-based person retrieval with uncertainty learning under noisy correspondence [64] and chat-driven interaction for person retrieval [65].

(3) Economic and reliability-domain metrics across Layers 1-6

These metrics translate system-level behavior into financial outcomes: TCO, cost-per-toke, TLOE, SLA penalties, redundancy cost, depreciation, downtime loss, grid price sensitivity, and reliability-adjusted lifecycle cost. They quantify how physical inefficiencies and computational behavior ultimately manifest as cost.

## V.3. Cross-Cell Interactions: Metric Propagation Across the Stack

A central feature of the unified taxonomy is that metrics propagate across cells, forming pathways rather than remaining static.

For example:

- Grid → Facility → Compute → Runtime → Economic

Carbon intensity influences cooling efficiency, which influences accelerator throttling and utilization, which influences training time and inference cost.

- Facility → Compute → Runtime

Thermal headroom affects GPU peak throughput, which affects distributed scaling efficiency.

- Network → Runtime → Economic

Collective-operation latency increases job completion time, raising cost per training step.

- Reliability → Economic

MTBF/MTTR affects required redundancy, which affects total TCO. These interactions can be represented formally using the Metric Propagation Graph introduced later in Section V.7.
Every directed edge in the MPG corresponds to a transformation operator mapping a metric in cell $(i, j)$to its impact in cell $(k, l)$.

Additionally, the unified taxonomy provides a principled basis for comparing heterogeneous systems of infrastructures that differ in technology, geography, workload type, or economic environment:

- NVIDIA DGX vs Google TPUv4/v5 clusters: Differences appear in (3,2), (4,2), and (6,3).
- Wind-powered Nordic data centers vs fossil-heavy grids: Differences appear strongly in (1,1), (1,3), (2,1), and downstream cost cells.
- Inference-optimized clusters vs training-optimized superpods: The largest differences occur in (4,2), (5,3).
- Cloud vs on-prem deployments: Cost structure differences reflect in (6,3) and (3,3).

Because all metrics are mapped into the same 6×3 coordinate system, comparison no longer requires ad-hoc assumptions or incompatible KPIs.

## V.4. Integration with Existing Metric Families

A significant insight from the taxonomy is that major existing metric families only occupy a small subset of the 18 cells:

- ISO/IEC 30134 xUE metrics → primarily (2,1) → Data Center teams optimize PUE
- MLPerf Training/Inference → primarily (5,2) → ML teams report MLPerf
- TCO / LCOE / LCOC → primarily (6,3) Finance teams model TCO
- Reliability engineering metrics (SLO/MTBF) → primarily (6,2) → Network teams optimize interconnect latency
- HPC Green500 FLOPs/W → primarily (3,2) → Finance teams model TCO

This reveals the fragmentation of current practice that none of these metrics families captures cross-layer interactions.
The unified taxonomy explicitly identifies these blind spots, offering a foundation for integrated assessment.

The unified 6×3 taxonomy serves as both a theoretical contribution and a practical analytical framework for AI infrastructure. It brings physical, computational, and economic metrics into a coherent structure that captures cross-layer dependencies and cost formation mechanisms.
By bridging historically isolated metric families, the taxonomy enables: (1) systematic economic reasoning; (2) Mechanism-based cost decomposition; (3) industry-standard benchmarking; (4) sustainability and carbon analysis; (5) infrastructure optimization.

# Section VI. The Metric Propagation Graph (MPG): A Topological Model for Systemwide Metric Dynamics

The unified 6×3 taxonomy developed in the previous section provides a complete structural map of the AI-infrastructure metrics universe. Yet the taxonomy itself is static: it specifies where metrics belong, but not how they interact, nor how physical, computational, or economic changes propagate across the system. Real-world AI infrastructure, however, is inherently dynamic. Grid fluctuations alter cooling requirements; thermal stress reduces accelerator efficiency; workload slowdowns inflate cost-per-training-step; and reliability events propagate through queues, networks, and service-level guarantees.

To capture these dependencies, I introduce the Metric Propagation Graph (MPG), a graph-theoretic model that elevates the taxonomy from a classification system to a systems-level representation of metric dynamics. MPG represents each metric as a node positioned in the 6×3 coordinate system and links them through directed edges encoding how changes in one metric influence others across layers and domains. This structure makes cascading effects analyzable, bottlenecks identifiable, and economic consequences traceable to their physical origins.

## VI.1 Concepts and Definition

A core motivation for MPG is to overcome a long-standing barrier in AI-infrastructure evaluation: the difficulty of building integrated or adjusted metrics (e.g., PUE-adjusted TCO, carbon-normalized throughput, reliability-adjusted cost-per-inference). Such metrics require knowing precisely how heterogeneous indicators, physical, computational, and economic, relate to each other. MPG provides this missing structure. By embedding all metrics in a unified topology with explicit directional relationships, MPG becomes a metric composability framework, enabling the derivation of coherent, semantically grounded composite metrics.

MPG captures propagation, the dominant mechanism by which physical or computational constraints manifest as economic or service-level outcomes.

Let the unified taxonomy define a set of nodes

$$V = \{(i, j) \mid i \in \{1, \dots, 6\}, j \in \{1,2,3\}\},$$

where $i$ indexes the infrastructure layer and $j$ the meta-domain. Each metric is treated as a variable $M_{i,j}$ associated with its cell. The MPG is a directed multigraph $G = (V, E)$ where each edge

$$e = (u \rightarrow v) \in E$$

represents a causal or functional dependency between metrics. Such dependencies arise from well-understood engineering pathways: thermal loads influence facility PUE that influences accelerator frequency; accelerator efficiency influences training throughput; runtime throughput influences cost per training token. The MPG makes these relationships explicit. Each edge is assigned a propagation operator

$$P_{u \rightarrow v}: M_u \mapsto M_v$$

that describes how a change in metric $M_u$ affects $M_v$, or integrates the metrics built based on $M_u$ with incorporating $M_v$. These operators may be linear, nonlinear, probabilistic, or time-delayed. By composing edge operators along a path, I obtain multi-step propagation chains that define how disturbances in one metric affect others across layers. In practice, propagation operators are obtained from physics models (e.g., cooling capacity curves), performance scaling laws (e.g., Amdahl's law, network latency models), or economic discounting functions. The complete mathematics of $P_{u \to v}$ (amplification factors, edge weights, and propagation compositions) is presented in Appendix B.

## VI.2. Practical Industry Value

A unified MPG provides three forms of value that existing metric families cannot deliver in isolation. First, it offers interpretability: the directed structure reveals bottlenecks, choke points, and leverage nodes, those with high in-degree, high out-degree, or strong propagation weights, allowing operators to understand how disturbances or improvements at one layer propagate across the system. Second, it enables metric synthesis: propagation chains naturally generate composite metrics such as PUE-adjusted cost per token, carbon-normalized throughput, and reliability-adjusted TCO, all of which are economically meaningful but cannot be defined within any single metric family. Third, it supports optimization: because tradeoffs among density, cooling, energy, carbon, reliability, and economics are encoded as graph relationships, MPG provides a principled structure for multi-objective optimization over cross-layer constraints. These capabilities collectively make MPG a practical, system-level economic tool rather than a purely diagnostic or monitoring artifact.

**(1) Systemwide Interpretation and Strength**

The MPG first builds up a complete map of essential metrics with their roles and positions in the ecosystem, their bound of function, and a statistics framework of field importance and industry hot zones, providing a unified interpretive lens for understanding AI infrastructure as a comparable system.

Cross-layer economic outcomes, such as cost-per-training-step or cost-per-inference, often arise from propagation chains that span multiple layers and domains. For example, a shift in grid carbon intensity (Layer 1, Domain 1) influences facility-level carbon conversion (Layer 2), alters accelerator thermal behavior (Layer 3), affects workload throughput (Layer 5), and ultimately changes economic cost structures (Layer 6, Domain 3). MPG makes such dependencies explicit, directional, and traceable.

The topology also reveals bottlenecks (nodes with high inbound degree where multiple inefficiencies converge) and leverage points (nodes with high outbound degree whose improvement yields large downstream gains). Additionally, MPG identifies amplification chains, where small perturbations in physical or computational metrics propagate nonlinearly into significant performance, reliability, or cost impacts. These capabilities transform otherwise fragmented KPIs into a coherent, causally interpretable system map.

**(2) Application Fields**

**MPG as a Generator of Integrated Metrics**

Because every metric is positioned in a directed dependency graph, the MPG defines not only where a metric sits, but how its influence propagates upward or downward across physical, computational, and economic layers. This structure makes it possible to derive integrated, adjusted, or augmented metrics in a mathematically coherent and interpretable way.

By positioning each metric within a directed dependency graph, MPG forms the structural foundation for a metric synthesis framework. For any propagation chain $A \rightarrow B \rightarrow C$, MPG defines the functional basis for constructing an integrated metric $M$that represents the aggregate effect of $A$on $C$. When multiple chains converge on an economic node, such as cost-per-training-step, carbon cost, or reliability-adjusted OpEx, the topology provides a natural decomposition structure for deriving adjusted or augmented metrics with weights grounded in the underlying propagation dynamics. For teams seeking composite indicators that balance energy, cooling, performance, reliability, and cost, the MPG effectively tells them which inputs matter, how they interact, and in what proportions.

This capability supports a broad family of next-generation infrastructure metrics, including:

- PUE-adjusted cost-per-training-step;
- carbon-normalized throughput;
- thermal-adjusted MLPerf efficiency;
- reliability-adjusted TCO;
- density-normalized cooling efficiency;
- failure-adjusted utilization.

Because these indicators derive directly from formally encoded propagation paths, they avoid the arbitrary weighting, boundary ambiguity, and semantic inconsistency that characterize ad hoc composite metrics today. MPG therefore establishes the foundation for a metric composability theory, a systematic, interpretable method for constructing cross-layer metrics that reflect systemwide economic and operational behavior.

**MPG as a Platform for Multi-Metric Optimization**

Infrastructure planning commonly involves countervailing metric directions: higher server density reduces amortized CapEx and land or water cost, but raises power density, cooling overhead, and thermal-induced failure sensitivity; workload consolidation improves utilization but increases correlated failure risk; low-carbon grid regions may incur latency or network-cost penalties.

MPG provides the graph structure needed to express these trade-offs rigorously. Countervailing metrics appear as nodes connected through weighted, signed edges that capture their directional influence. This enables fault-aware, reliability-aware, carbon-aware, and density-aware optimization to be formulated directly within the MPG topology. It also supports the derivation of optimization-oriented composite indicators such as density-adjusted cooling cost, failure-adjusted throughput, and carbon-aware marginal cost of compute.

Thus, MPG converts multi-objective trade-off management from a heuristic exercise into a structured, graph-driven optimization problem.

**MPG as a Standardization Mechanism**

Metric boundaries in AI infrastructure remain inconsistent across organizations: cooling overhead may or may not be counted toward PUE; “utilization” may refer to SM occupancy, scheduling efficiency, or model-level throughput; carbon attribution varies across facility, procurement, and execution layers. This fragmentation impedes cross-team communication, benchmarking, and investment decisions.

MPG provides a formal boundary-normalization mechanism. Each metric occupies a specific layer-domain coordinate and a directed dependency context, which jointly define its operational scope, upstream inputs, and downstream influence. This canonical positioning eliminates semantic drift and resolves boundary ambiguity by construction.

This structured topology creates a neutral semantic backbone for all stakeholders:

- OEMs can position hardware-level efficiency, memory behavior, and thermal envelopes within a shared dependency graph linking component behavior to system-level outcomes.
- Cloud providers gain a standardized substrate for interpreting facility metrics, accelerator performance, reliability indicators, and carbon signals across heterogeneous regions.
- Data-center operators can map power, cooling, density, and fault-domain metrics to workload performance and cost outcomes with consistent definitions.
- Model developers gain clear visibility into how workload-level behavior (e.g., batch size, convergence time) propagates into energy, carbon, and TCO consequences.

By enabling cross-team comparability, shared dashboards, and consistent evaluation frameworks, MPG establishes the common substrate upon which integrated metrics, optimization policies, and carbon-reliability-cost tradeoff analysis can be built. It transforms currently fragmented measurement practices into a coherent system-level foundation for decision-making.

# VII. System-Level Bottlenecks and Future Directions

The unified 6×3 metric taxonomy and the Metric Propagation Graph developed in this paper reveal that future bottlenecks will no longer be dominated by isolated components, such as compute throughput or FLOP capacity, but by multi-layer interactions shaped by energy, thermal, network, and operational dynamics. This section outlines three frontier directions that, taken together, define the next decade of AI-infrastructure research.

(1) Power, Thermal, and Siting Constraints Become the Dominant Limiters of AI Growth

While public discourse continues to center on GPU shortages, the binding constraint on AI expansion is rapidly shifting toward power availability, thermal feasibility, and grid compatibility. Accelerator TDPs approaching 1 kW and rack envelopes exceeding 100-150 kW place most existing facilities outside their physical design limits. As a result, even well-provisioned clusters face derating, throttling, and capacity fragmentation.

Future research must integrate power-aware, carbon-aware, and thermally-aware workload orchestration, linking TTT, PUE/ERE, LMP, marginal emissions factors, and real-time temperature limits into unified scheduling models. These developments need to extend beyond the facility boundary: grid-side analytics, renewable availability, congestion, locational marginal pricing, and microgrid coupling, must be treated as co-optimizers rather than exogenous inputs. AI infrastructure will increasingly operate within a capacity-constrained, carbon-regulated, and geographically differentiated power system, requiring co-designed strategies for siting, load shifting, and energy procurement.

(2) Distributed Training Efficiency Will Depend More on Networks Than on FLOPs

As models cross trillion-parameter scales and mixture-of-experts architectures proliferate, communication, not computation, has become the binding constraint. Metrics such as all-reduce latency, bisection bandwidth, oversubscription ratio, topology depth, and HBM memory coupling now dominate realized FLOP utilization across all major workloads.

Yet current benchmarks often abstract away the network as a fixed penalty, failing to account for its decisive role in governing scaling efficiency. The next frontier requires:

- network-aware scaling laws extending Amdahl and roofline models to fabric congestion;
- graph-topology optimization frameworks for jointly designing routing, sharding, and parallelism strategies;
- co-designed compute-memory-network architectures where model parallelism, kernel fusion, and pipeline schedules adapt dynamically to interconnect conditions.

For clusters operating at thousand-million GPU scales, network performance will increasingly dictate training cost, job preemption behavior, and overall TCO.

### (3) Data Pipeline Performance and Lifecycle Economics Become Key Research Problems

Despite the focus on GPU performance, large-scale training runs frequently remain I/O-bound or scheduler-bound, with accelerators idling due to insufficient data throughput, dataset serialization inefficiencies, input-pipeline stalls, or job fragmentation. This class of bottleneck directly impacts FLOP realization, TTT, and per-step efficiency, yet remains underrepresented in most academic and industrial analyses.

In parallel, the lifecycle economics of AI systems, driven by energy prices, carbon markets, water scarcity, depreciation cycles, reliability risk, and supply-chain constraints, are becoming as important as hardware capabilities. Open research directions include:

- modeling TCO/GPU-hour under uncertainty and dynamic utilization;
- optimizing training under carbon, water, and thermal budgets;
- quantifying value degradation due to thermal derating, network-induced inefficiency, or hardware-aging;
- developing multi-objective optimization frameworks that integrate performance, energy, carbon, reliability, and long-run cost into a single decision space [53].

Achieving sustainable AI growth will require re-framing infrastructure economics not as an afterthought but as a central design axis.

Figure 1 highlights a critical structural insight: removing one bottleneck simply shifts pressure to another, because the constraints governing AI systems are inherently cross-layer. Reducing thermal overhead without addressing rack power density increases throttling; improving FLOP efficiency without solving I/O starvation produces no net speedup; expanding cluster scale without improving all-reduce latency disproportionately increases TCO. This interdependency, made explicit in the Metric Propagation Graph, underscores why traditional single-metric optimization has reached its limit. A complete list of representative metrics associated with each bottleneck appears in Appendix F.

Figure 1. Map of Key AI Infrastructure Bottlenecks by Metrics

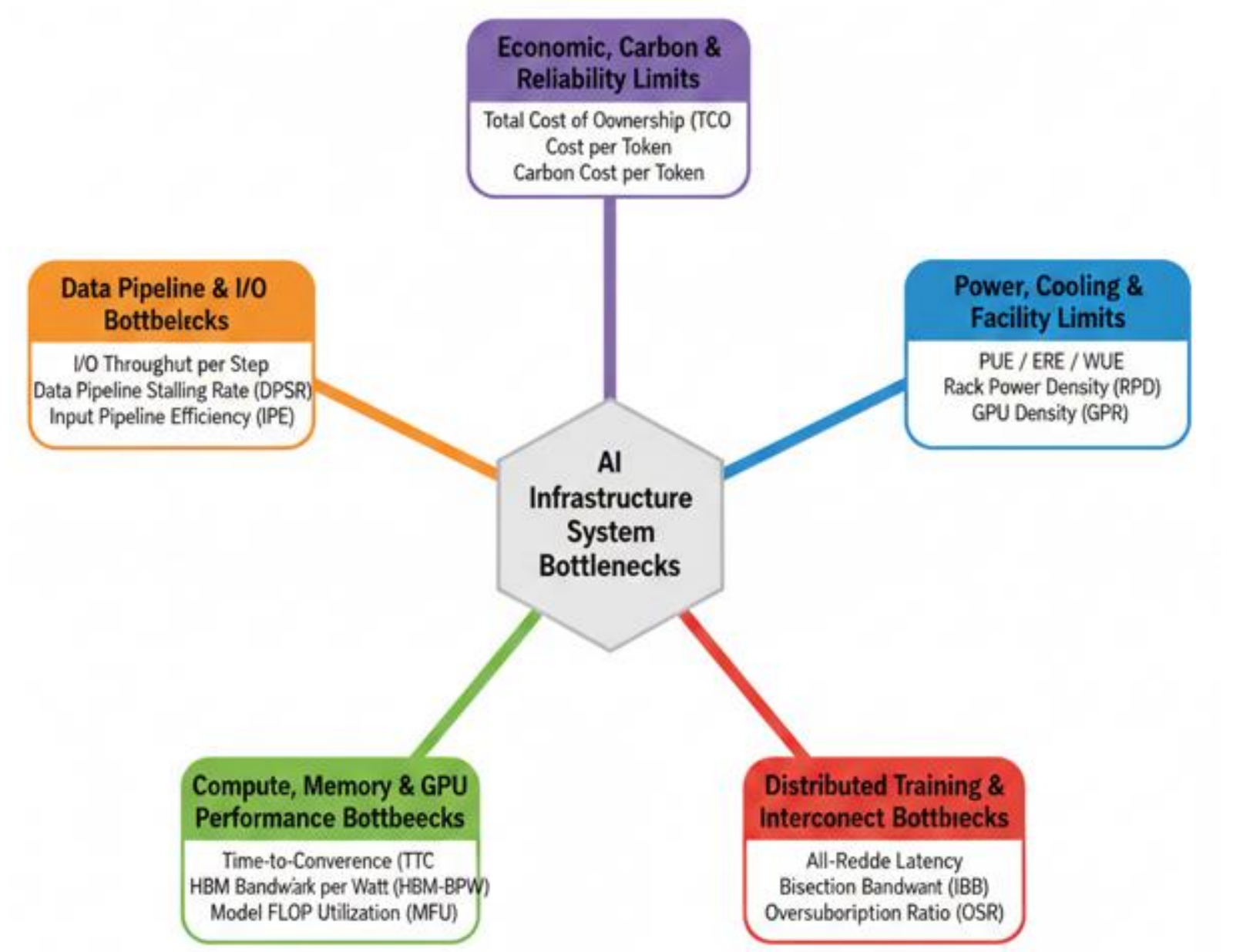

*Notes*: Figure 1 reveals that AI infrastructure bottlenecks do not arise from isolated components but from the coupled limits of data movement, compute efficiency, thermal and electrical constraints, distributed synchronization, and lifecycle economics. Each spoke highlights a structurally distinct failure mode, data stalls, GPU underutilization, thermal headroom collapse, interconnect saturation, and escalating cost-per-token, that together define the systemwide frontier of AI scalability.

# Conclusions

AI systems are increasingly defined by the capabilities and limitations of the infrastructure that supports them. Yet the metrics used to evaluate modern AI infrastructure, spanning physical efficiency, computational behavior, and lifecycle economics, have historically evolved in isolation, resulting in fragmented assessment practices and inconsistent decision frameworks. This paper addresses this gap by developing a unified, structurally grounded framework that integrates these perspectives into a coherent and interpretable system.

I first established a three-meta-domain decomposition that captures the semantic foundations of all major infrastructure metrics: physical facility efficiency, compute and workload efficiency, and economic and reliability efficiency. I then introduced a six-layer architectural hierarchy, from the power grid through facility, hardware, interconnect, and runtime to service economics, that characterizes the physical pathways through which performance, energy, reliability, and cost propagate. Together, these components form a 6×3 unified taxonomy that systematically maps the metric landscape and clarifies the domain and structural locus of each metric.

Building on this foundation, I proposed the MPG, a graph-structured formulation that expresses cross-layer dependencies and provides a principled mechanism for systemwide interpretation. The MPG enables the synthesis of composite metrics, reveals leverage points and bottlenecks, and supports multi-objective optimization across density, cooling, carbon, reliability, and operational cost. This representation makes explicit the causal pathways often implicit in infrastructure analysis and bridges the longstanding divide between engineering heuristics, computational benchmarks, and lifecycle economic models.

The unified framework developed here has implications beyond the analysis presented in this paper. It provides a transparent basis for cross-site comparison, supports reproducible benchmarking across vendors and architectures, and offers a foundational structure for integrating energy-system constraints and carbon policy into AI-infrastructure planning. As AI adoption continues to expand and infrastructure demands approach the limits of regional grids and cooling capacity, the need for consistent, interpretable, and scientifically grounded metrics will only intensify.

Future research may extend this framework in several directions. One avenue is to incorporate stochastic models of workload variability and grid uncertainty into MPG-based optimization. Another is to align the taxonomy with emerging hardware abstractions, such as disaggregated memory or photonic interconnects, which may reshape metric propagation pathways. Finally, the framework can support the development of standardized reporting guidelines for AI infrastructure, analogous to established standards in energy systems, which could enable more transparent industry comparisons and inform both private-sector investment and public policy.

By unifying the physical, computational, and economic layers of AI infrastructure, this work provides a structural foundation for the next generation of infrastructure research, design, and governance. As the scale and societal impact of AI continue to grow, such unified and interpretable frameworks will be essential for guiding sustainable, reliable, and economically coherent development.

# Appendix A. Comprehensive Metrics Dictionary for AI Infrastructure Analysis

This appendix provides a unified and systematic dictionary of the metrics used throughout the paper's six-layer, cross-domain AI infrastructure framework. Because the full stack of AI systems spans physical grid interactions, facility-level thermodynamics, computing hardware, networking fabrics, machine-learning execution, service-level reliability, and economic performance, the terminology used in industry and academic literature is highly heterogeneous. Metrics are often defined inconsistently across organizations, including hyperscalers, chip vendors, ML benchmarking groups, and power-systems agencies, creating ambiguity in cross-layer economic or efficiency analysis.

To resolve this fragmentation, Appendix A consolidates all metrics referenced across the 18 analytical cells (6 Layers × 3 Domains) into a single, coherent reference. The appendix includes all metrics.

Table A. Unified Metrics Dictionary Across All Layers and Domains of AI Infrastructure

| | Metrics | Definition | Interpretation |
|---|---|---|---|
| **Grid & Sustainability Layer - Domain 1: Physical Efficiency** | Marginal Emissions Factor (MEF) | $MEF = \frac{\Delta Emissions}{\Delta Load}$ | The marginal emissions factor measures the change in grid-level $CO_2$ emissions from a marginal increase or decrease in electrical load. It reflects the carbon intensity of the marginal generator. |
| | Time-Varying Grid Carbon Intensity | $CI(t) = \frac{E_{CO_2}(t)}{P_{grid}(t)}$ | Carbon intensity represents the instantaneous emission rate of the grid at time $t$, generally expressed in $gCO_2/kWh$. It captures temporal fluctuations in the grid's generation mix. |
| | Renewable Availability | $RA(t) = \frac{P_{renewable}(t)}{P_{total}(t)}$ | Share of electricity generated from renewable sources (solar, wind, hydro, etc.) at time $t$. Indicates opportunities for renewable-aligned workload shifting. |
| **Grid & Sustainability Layer - Domain 2: Compute & Workload Efficiency** | Grid-Constrained Workload Shiftability | Key metric: Shiftable Workload Fraction (SWF) $SWF = \frac{W_{shiftable}}{W_{total}}$ | Fraction of workload that can be temporally shifted to align with low-carbon or low-cost periods. Measures the workload's temporal flexibility under grid constraints. |
| | Carbon-Aware Scheduling Limits | Key metric: Emissions-Reduction Potential (ERP) $ERP = \frac{E_{baseline} - E_{CAS}}{E_{baseline}}$ | The maximum proportion of emissions that can be reduced by shifting execution into lower-carbon intervals, assuming perfect knowledge of carbon intensity signals. |
| | Workload Flexibility Under Grid Conditions | $WF(t) = \frac{W_{executed\ in\ flexible\ window}(t)}{W_{total}(t)}$ | Represents the share of active workload that responds to real-time grid conditions (curtailment, congestion, or carbon signals). Useful for carbon-aware or demand-response compute. |
| **Grid & Sustainability Layer - Domain 3: Lifecycle Economics & Reliability Risk** | Time-Varying Electricity Price (LMP / RTO Pricing) | $Price(t) = \lambda_{LMP}(t)$ | The locational marginal price reflects the marginal cost of serving the next unit of electricity at time $t$, incorporating energy, congestion, and loss components. |
| | Carbon Pricing Exposure | $CPE(t) = CI(t) \times CarbonPrice(t)$ | Effective monetary exposure to carbon cost at time $t$. Used when carbon pricing, carbon taxes, or cap-and-trade systems apply. |
| | Carbon Cost per Token | $C_{\text{carbon,token}} = \frac{gCO_2}{\text{token}} \cdot \text{carbon price}$ | Monetary cost of carbon emissions attributable to generating one unit of AI output (one token), integrating facility efficiency, grid carbon intensity, and carbon pricing. It reflects how sustainability constraints propagate into model inference cost. |
| | Location-Based Energy Arbitrage Potential | $AP = max_{\ i,j}(Price_j)$ | Measures the economic advantage of running compute in lower-price regions and shifting away from higher-price regions. Relevant in multi-region data center portfolios. |
| | Grid Interaction Cost | $C_{interaction} = C_{energy} + C_{demand} + C_{carbon} + C_{DR\ penalties}$ | Composite cost of interacting with the grid, including energy, demand, carbon charges, and demand-response penalties. Captures full economic exposure to grid-side fluctuations. |
| **Facility Layer - Domain 1: Physical Performance & Efficiency** | GPUs per Rack (GPR) Formula | $\text{GPR} = \frac{N_{\text{GPU}}}{N_{\text{racks}}}$ | The number of GPUs that can be installed in a single rack, constrained by power, cooling, and network topology. A fundamental limit for cluster design and TCO. |
| | Rack Power Density (RPD) | $\text{RPD} = \frac{P_{\text{rack}}}{\text{rack footprint}}$ | Measures the maximum deliverable power envelope for a rack and is the primary physical constraint on deploying GPU-dense clusters. High RPD is required for modern AI clusters (30-120 kW/rack). |
| | Carbon Price Exposure | $CPE = CUE \times CarbonPrice$ | Monetary exposure to carbon pricing mechanisms based on emissions-related energy use. |
| | Cooling System COP (Coefficient of Performance) | $COP = \frac{Q_{cooling}}{P_{input}}$ | Thermodynamic efficiency of cooling systems; higher COP indicates more efficient cooling. |
| | Cooling Power Fraction | $CPF = \frac{P_{cooling}}{P_{facility}}$ | Fraction of total facility power dedicated to cooling operations. |
| | Thermal Design Margin (Cooling Headroom) | $TDM = T_{throttle} - T_{supply}$ | Thermal buffer between cooling supply temperature and throttling threshold. |
| **Facility Layer - Domain 2: Compute & Workload Efficiency** | Latency Penalty During Facility Events | $LPF = \frac{L_{event} - L_{nominal}}{L_{nominal}}$ | Relative increase in latency due to cooling or power constraints. |
| | Throughput Degradation Factor (TDF) | $TDF = \frac{T_{nominal} - T_{event}}{T_{nominal}}$ | Fractional throughput loss caused by facility congestion or curtailment. |
| | Thermal-Throttling Duty Cycle (TTDC) | $TTDC = \frac{t_{throttled}}{t_{total}}$ | Portion of time hardware operates in throttled state due to thermal constraints. |
| | Workload Curtailment Ratio (WCR) | $WCR = \frac{W_{curtailed}}{W_{scheduled}}$ | Workload volume curtailed due to facility-level constraints. |
| | Facility-to-Compute Thermal Coupling Coefficient | $\kappa_{FC} = \frac{\partial T_{chip}}{\partial T_{supply}}$ | Sensitivity of chip temperature to changes in cooling supply temperature. |

| **Facility Layer - Domain 3: Lifecycle Economics & Reliability Risk** | Energy Opex Share | $EOS = \frac{C_{energy}}{C_{total\ facility\ opex}}$ | Share of facility operating cost driven by electricity consumption. |
|---|---|---|---|
| | Cooling Opex Share | $COS = \frac{C_{cooling}}{C_{total\ facility\ opex}}$ | Share of opex attributable to cooling infrastructure and processes. |
| | Demand-Response Revenue Potential | $DRP = \sum_{t} (P_{incentive}(t) \times \Delta L(t))$ | Revenue achievable through participation in grid demand-response events. |
| | Carbon Cost (from CUE × Carbon Price) | $C_{carbon} = CUE \times CarbonPrice$ | Carbon cost incurred due to emissions-related energy usage. |
| | Reliability Downtime Cost | $C_{downtime} = \lambda_{failure} \times L_{downtime}$ | Expected annual cost of downtime from facility system failures. |
| **Compute Hardware Layer - Domain 1: Physical Performance & Efficiency** | Device Power Draw | $P_{device} = V \times I$ | Instantaneous electrical power consumption of a compute device; key determinant of thermal load and energy requirements. |
| | BM Bandwidth per Watt (HBM-BPW) | $\text{HBM-BPW} = \frac{BW_{\text{HBM}}}{P_{\text{memory}}}$ | Measures the energy efficiency of on-package high-bandwidth memory (HBM). Critical for LLM training workloads where memory bandwidth, not FLOPs, is often the bottleneck. |
| | Temperature-Performance Coupling | $\gamma_{TP} = \frac{\partial Perf}{\partial T_{chip}}$ | Sensitivity of device performance to chip temperature; captures thermal throttling behavior. |
| | Thermal Headroom | $TH = T_{throttle} - T_{chip}$ | Temperature margin before throttling begins; indicates thermal safety and cooling adequacy. |
| | Voltage-Frequency Scaling (VFS) Efficiency | $\eta_{VFS}(f) = \frac{Perf(f)}{P(f)}$ | Energy efficiency of compute under different clock frequencies and voltages; used in power-capping and DVFS policies. |
| | Idle Power Fraction | $IPF = \frac{P_{idle}}{P_{TDP}}$ | Fraction of TDP consumed during idle state; reflects static inefficiency of hardware when underutilized. |
| **Compute Hardware Layer - Domain 2: Compute & Workload Efficiency** | FLOPs per Watt (FLOPs/W) | $\frac{FLOPs}{P}$ | Standard compute-efficiency metric representing computational throughput per watt of power. |
| | Joules per Operation | $E_{op} = \frac{E_{total}}{N_{operations}}$ | Energy required per operation; reciprocal of FLOPs/W. |
| | Accelerator Utilization | $U = \frac{t_{active}}{t_{total}}$ | Fraction of time an accelerator is actively executing workloads; low utilization indicates scheduling or pipeline bottlenecks. |
| | Tokens per Joule (LLM Efficiency) | $\frac{Tokens}{E}$ | Measures energy efficiency of LLM inference, increasingly used as a standard efficiency metric for transformer models. |
| | Memory Bandwidth Utilization | $MBU = \frac{B_{used}}{B_{peak}}$ | Fraction of available memory bandwidth consumed; captures memory-bound bottlenecks. |
| **Compute Hardware Layer - Domain 3: Lifecycle Economics & Reliability Risk** | Hardware TCO Contribution | $TCO_{HW} = C_{capex} + \sum_{t} \frac{C_{opex}(t)}{(r)^t}$ | Portion of total cost of ownership attributable to compute hardware (purchase, maintenance, energy, depreciation). |
| | Memory Power Share (MPS) | $\text{MPS} = \frac{P_{\text{memory}}}{P_{\text{system}}}$ | Fraction of total system power consumed by DRAM/HBM. Often 25-40% in modern AI servers (per "AI Tax" paper). Key for cooling, provisioning, and TCO-per-token. |
| | Depreciation Curve | $V(t) = V_0(1-d)^t$ | Hardware value after $t$periods under geometric depreciation rate $d$. |
| | Cost per Accelerator-Hour | $C_{acc-hour} = \frac{C_{capex}/L + C_{opex}}{H_{annual}}$ | Effective hourly cost for running one accelerator, inclusive of amortization and operating expenses. |
| | Failure-Driven Replacement Cost | $C_{rep} = \lambda_{fail} \times C_{unit}$ | Expected annual replacement cost for failed compute devices; key in reliability-aware economic modeling. |
| | Hardware Lifetime Efficiency | $HLE = \frac{Useful\ Compute\ Output}{Lifetime\ Cost}$ | Overall efficiency of hardware over its usable life, defined as compute delivered per dollar spent. |
| **Networking & Interconnect Layer - Domain 1: Physical Performance & Efficiency** | Network Power Overhead | $NPO = P_{network} = P_{switches} + P_{NICs} + P_{optics}$ | Total power consumed by network infrastructure (switches, NICs, optics). Indicates physical overhead of interconnect relative to compute. |
| | Network Diameter / Hops | $D = max\ shortest - path\ length\ between\ nodes$ | Network Diameter is the maximum shortest path (in hops) between any two accelerators in the cluster topology. Fatal to all-reduce, MoE routing, parameter sharding time lag. LLM inference + MoE routing = latency highly sensitive to network diameter |
| | Switch Load Power Ratio | $SLPR = \frac{P_{switch}}{P_{rack}}$ | Fraction of rack-level power used by switching hardware; reflects network intensity of a cluster. |
| | Rack/Pod Thermal Load from Networking | $TL_{net} = \frac{P_{network}}{P_{total}}$ | Share of thermal load contributed by networking equipment; influences cooling planning and hot-aisle design. |
| | Interconnect Energy per Bit | $E_{bit} = \frac{E_{network}}{B_{transmitted}}$ | Energy required to transmit one bit over the fabric; key for large-scale distributed training. |
| **Networking & Interconnect Layer - Domain 2: Compute & Workload** | Pipeline Efficiency (PE) | $PE = \frac{T_{ideal}}{T_{actual}}$ | Measures how efficiently a distributed pipeline (training or inference) achieves ideal end-to-end throughput. Captures pipeline bubbles and imbalance. |
| | Bisection Bandwidth （BB/sec） | $\text{BB/sec} = \frac{\text{IBB}}{2}\text{(cluster evenly partitioned)} = \min_{\text{cluster bisection}} \sum_{\ell} BW_{\ell}$ | Classical HPC bandwidth metric used to characterize communication capacity under worst-case workload partitioning. High BB is essential for large-model all-reduce and for maintaining low synchronization overhead in transformer training. |
| | Interconnect Bisection Bandwidth (IBB) | $\text{IBB} = \min_{\text{cut}} \sum_{\text{links crossing cut}} \text{bandwidth}$ | Measures the minimum bandwidth available across any cluster-wide network bisection and is the dominant determinant of distributed training scalability. Essential for superpods design. |
| | Bandwidth per Watt | $BPW = \frac{B}{P_{network}}$ | Data-movement throughput per watt of networking power. Higher values indicate more energy-efficient interconnect hardware. |
| | All-Reduce Latency (ARL) | $L_{collective} = = f(topology, message\ size, congestion)$ | Latency of collective communication (AllReduce, Broadcast, etc.), central to distributed deep learning performance. |
| | Congestion Penalty (CP) | $CP = \frac{L_{p99} - L_{p50}}{L_{p50}}$ | Quantifies tail-latency degradation due to network congestion; reflects imbalance and queueing delays under heavy load. |
| | Topology-Induced Throughput Loss | $TTL = 1 - \frac{T_{observed}}{T_{full-bisection}}$ | Fraction of throughput lost because real-world topology deviates from ideal full-bisection bandwidth (e.g., oversubscription). |
| | Effective Bandwidth Utilization | $EBU = \frac{B_{useful}}{B_{peak}}$ | Efficiency of available network bandwidth actually used for productive compute-related communication. |
| **Networking & Interconnect Layer - Domain 3: Lifecycle Economics & Reliability Risk** | Interconnect CapEx Contribution | $C_{net,capex} = C_{switches} + C_{optics} + C_{NICs}$ | Capital cost associated with networking hardware as part of a cluster or data center build-out. |
| | Network Scaling Cost | $C_{scale} == f(N, fabric\ type, oversubscription\ ratio)$ | Total cost of scaling the network as cluster size $N$grows; reflects topology, bisection needs, and redundancy requirements. |
| | Topology-Driven Efficiency Loss | $TDL = \frac{T_{ideal} - T_{topology}}{T_{ideal}}$ | Portion of throughput lost due to non-ideal network topology design (e.g., fat-tree oversubscription). |
| | Network Failure Rate | $\lambda_{net} = \frac{Number\ of\ failures}{Time}$ | Frequency of network-component failures (optics, NICs, switches); affects reliability planning and redundancy needs. |
| | Network Reliability Cost | $C_{reliability} = \lambda_{net} \times L_{event}$ | Expected economic cost associated with network outages or degraded service. |
| **ML Execution Layer - Domain 1: Physical Performance & Efficiency** | Thermal-Induced Failure Rate | $\lambda_{thermal} = f(T_{chip}, t_{exposure})$ | Failure rate caused by prolonged exposure to elevated chip temperatures. Often modeled using Arrhenius acceleration factors. |
| | Power-Quality-Induced Failure Rate | $\lambda_{PQ} = f(voltage\ sag, harmonics, transients)$ | Failure frequency arising from unstable or degraded power quality, including brownouts and voltage spikes. |
| | Compiler-Induced Energy Overhead | $CEO = \frac{E_{compiled} - E_{ideal}}{E_{ideal}}$ | Additional energy consumption caused by suboptimal compiler or kernel scheduling decisions. |
| | Idle Energy from Inefficient Scheduling | $E_{idle} = P_{idle} \times t_{idle}$ | Total idle energy burned due to scheduling stalls, synchronization delays, or uneven pipeline workloads. |

| ML Execution Layer - Domain 2: Compute & Workload Efficiency | MLPerf Time-to-Train (TTT) | $TTT = t_{converged} - t_{start}$ | Time required for a model to reach a defined target accuracy. Primary metric for MLPerf Training benchmark. |
|---|---|---|---|
| | MLPerf Throughput (QPS/TPS) | $TPS = \frac{Total\ Queries}{t_{window}}$ | Inference throughput under MLPerf constraints; widely used industry benchmark. |
| | Data Pipeline Stalling Rate (DPSR) | $DPSR = \frac{\text{Stalled Steps}}{\text{Total Steps}}$ | Fraction of training steps where GPUs stall waiting for input data (decode, augment, batch assemble, prefetch). |
| | Job Packing Efficiency/Fragmentation Loss | $Fragmentation\ Loss = 1 - JPE$ | Utilization loss not caused by compute or data bottlenecks, but caused by orchestration geometry |
| | Effective Model Parallel Efficiency (MPE) | $MPE = \frac{\text{Achieved Throughput}}{\text{Ideal Parallel Throughput}}$ | It captures communication overhead, pipeline bubbles, imbalance among stages, and tensor-parallel sharding cost. |
| | Tokens per Second (LLM Serving Throughput) | $TPS_{LLM} = \frac{Tokens}{t}$ | Standard measure of LLM decoding throughput. |
| | Storage I/O Throughput per Training Step | $\text{IO}_{\text{per step}} = \frac{V_{\text{IO}}}{N_{\text{steps}}}$ | Measures dataset read bandwidth and I/O system ability to feed GPUs timely. GPU idle waiting for next batch. |
| | Pipeline Parallelism Efficiency | $PPE = \frac{T_{ideal}}{T_{pipeline}}$ | Captures pipeline bubbles and per-stage imbalance in multi-stage distributed execution. |
| | Energy per Checkpoint (EPC) | $EPC = \frac{E_{checkpoint}}{N_{checkpoints}}$ | Energy consumed per model checkpoint during training. |
| ML Execution Layer - Domain 3: Lifecycle Economics & Reliability Risk | Cost per Training (TPT) | $C_{step} = \frac{C_{total}}{N_{steps}}$ | Total cost (compute, energy, depreciation) amortized per training step. |
| | Cost per Token (Inference) | $C_{token} = \frac{C_{inference}}{N_{tokens}}$ or $C_{token} = \frac{P_{GPU} t_{lat}}{tokens}$ | Monetary cost of generating one token during inference. |
| | Marginal Serving Cost (MSC) | $MSC = \frac{\partial C}{\partial Q}$ | Additional cost of serving one more unit of workload (e.g., per 1k tokens or per request). |
| | Autoscaling Overhead (ASO) | $ASO = \frac{C_{autoscale} - C_{ideal}}{C_{ideal}}$ | Cost penalty due to autoscaling inefficiencies such as cold starts or idle buffer capacity. |
| | Expected Loss from Training Instability | $EL_{inst} = P_{diverge} \times C_{restart}$ | Expected loss from training runs that diverge and must be restarted. |
| | Annual Loss Expectancy (ALE) from Failures | $ALE = \lambda_{failure} \times L_{event}$ | Expected annual economic loss resulting from system failures (hardware, network, software). |
| Service Layer - Domain 1: Physical Performance & Efficiency | Redundancy Energy Overhead | $REO = \frac{P_{redundant}}{P_{total}}$ | Fraction of energy spent on redundant components (e.g., N+1, 2N systems) to ensure service reliability. |
| | Energy from Thermal Penalties (Derating Overhead) | $DEO = \frac{E_{derated} - E_{optimal}}{E_{optimal}}$ | Extra energy incurred when compute systems operate below optimal efficiency due to high temperatures or thermal derating. |
| | Service Energy Inefficiency Factor | $SEIF = \frac{E_{service}}{E_{compute}}$ | Overhead energy consumption required for achieving a given level of service quality (e.g., replication, retries). |
| | Thermal Stability Index | $TSI = \frac{T_{max} - T_{service}}{T_{max} - T_{nominal}}$ | Normalized indication of thermal stability of service nodes relative to safe operating bounds. |
| Service Layer - Domain 2: Compute & Reliability (SLO/SLA Metrics) | Service Level Indicator (SLI) | $SLI = \frac{Number\ of\ successful\ requests}{Total\ requests}$ | Base measurement of service correctness, availability, or latency adherence. |
| | Availability (A) | $A = \frac{Uptime}{Uptime + Downtime}$ | Standard measurement of system availability; often used in “five nines” reliability targets. |
| | SLO Violation Rate | $ViolationRate = \frac{Violations}{Total\ Requests}$ | Percentage of requests failing to meet defined service-level objectives (latency, throughput, correctness). |
| | Error Budget Consumption (EBC) | $EBC = \frac{SLO - SLI}{1 - SLO}$ | Fraction of allowable error budget consumed; widely used in SRE practice. |
| | Mean Time Between Failures (MTBF) | $MTBF = \frac{Total\ Operating\ Time}{Number\ of\ Failures}$ | Frequency metric describing how often service-impacting failures occur. |
| | Mean Time to Recovery (MTTR) | $MTTR = \frac{\sum Repair\ Time}{Number\ of\ Failures}$ | Average time to restore service after an outage. |
| | QoS Compliance Rate | $QCR = 1 - QDF = 1 - \frac{t_{QoS\ violated}}{t_{total}}$ | Fraction of time service meets required QoS/SLO standards. |
| Service Layer - Domain 3: Lifecycle Economics & Reliability Risk | Total Service Cost (TSC) | $TSC = C_{compute} + C_{network} + C_{energy} + C_{SRE} + C_{failures}$ | Full economic cost of providing service reliability and performance, including SRE labor, hardware, network, and failure losses. |
| | GPU Interconnect Oversubscription Ratio | $OSR = \frac{\text{GPU aggregate BW}}{\text{switch uplink BW}}$ | Measures how heavily the GPU-to-switch or GPU-to-GPU network bandwidth is oversubscribed relative to the aggregate bandwidth required by training workloads. |
| | SLA Penalty Cost | $C_{SLA} = \sum_i (L_i \times P_{penalty})$ | Cost incurred when service availability breaches contractual SLAs. |
| | Cluster Utilization Rate (CUR) | $\boldsymbol{CUR} = \frac{\text{Active Compute Time}}{\text{Total Provisioned GPU Time}}$ | Measures the fraction of total available accelerator capacity actively performing useful computation over time, determining GPU amortization efficiency. |
| | Token Throughput per Watt (TPW) | $\boldsymbol{TPW} = \frac{\text{Tokens Processed}}{\text{Energy (W\textbackslash{}cdotps)}}$ | Energy efficiency metric for inference (and sometimes training). Fits better for LLM inference than FLOP/W. Key for inference scaling, cost-per-token, and carbon footprint. |
| | Expected Loss from MTBF/MTTR Mismatch (ALE) | $ALE = \lambda_{failure} \times L_{event}$ | Expected annual financial loss due to reliability shortfalls (hardware, network, or service components). |
| | Operational Cost Elasticity | $OCE = \frac{\partial C_{service}}{\partial Q}$ | Additional service cost incurred per incremental unit of workload (queries, tokens, users). |
| | Service Lifetime Value (SLV) | $SLV = \sum_{t=1}^{T} \frac{Revenue(t) - Cost(t)}{(r)^t}$ | Net present value of operating a service over its lifetime. |

*Notes:* This table consolidates all metrics defined across 6x3 taxonomy into a single, unified reference spanning the full six-layer, three-domain analytical framework. Metrics include physical, thermal, compute-level, networking, ML execution, reliability, and economic indicators used throughout the paper’s propagation-graph and cross-layer analysis. Each entry preserves the original terminology used in Table 3 while providing a standardized mathematical formulation and concise interpretation for reproducibility. Metrics belonging to the XUE family (e.g., PUE, WUE, CUE, ERE) and TCO-related measures are omitted here, as they are documented separately in dedicated appendices.

# Appendix B. Formal Mathematical Specification of the Metric Propagation Graph (MPG)

This appendix provides the formal mathematical details underlying the Metric Propagation Graph (MPG) introduced in Section VI. The main text presents the conceptual foundation of the MPG; here I rigorously define node semantics, edge operators, propagation matrices, path composition rules, and conditions for amplification, attenuation, and economic convergence.

## B.1. Framework

**(1) Node Space and Metric Variables**

The unified taxonomy defines a finite node space

$$V = \{(i,j) \mid i \in \{1, \ldots, 6\},\ j \in \{1,2,3\}\},$$

where $i$ denotes the infrastructure layer (L1 to L6), $j$ denotes the meta-domain (D1 to D3). Each node is associated with a metric variable

$$M_{i,j}(t) \in R,$$

representing the instantaneous value of the metric at time $t$. The metrics may be physical (e.g., PUE), computational (e.g., FLOPs/W), or economic (e.g., cost per training token). Metrics may be scalar, vector-valued, or stochastic. For generality, I define:

$$M_{i,j}(t) \in R^k, k \geq 1.$$

**(2) Graph Structure**

The Metric Propagation Graph is a directed multigraph

$$G = (V, E),$$

where each edge

$$e = (u \to v) \in E$$

encodes a causal or functional dependency from metric $M_u$ to metric $M_v$. For nodes $u = (i,j)$ and $v = (k,l)$, I write

$$u \prec v$$

if $u$ is a causal predecessor of $v$. A key design constraint is acyclicity across layers:

$$if\ i < k, then\ paths\ u \to v \to \cdots \to u\ are\ prohibited.$$

Within a layer, lateral edges are permitted if supported by system behavior (e.g., physical ↔ economic coupling within Layer 6).

**(3) Propagation Operators**

Each edge $u \to v$ is associated with a propagation operator

$$P_{u \to v}: R^{k_u} \to R^{k_v},$$

mapping

$$M_v(t+1) = M_v(t)\ \oplus\ P_{u \to v}(M_u(t)),$$

where $\oplus$ is a node-specific aggregation operator (defined in A.4).

I identify five canonical operator classes, each corresponding to widely observed AI infrastructure propagation behaviors.

**Linear Propagation**

Models where changes propagate proportionally:

$$P_{u \to v}(M_u) = \alpha_{uv} M_u.$$

Common for thermal → performance effects, or performance → cost transitions.

**Multiplicative or Scaling Propagation**

$$P_{u\to v}(M_u) = M_v \odot (1 + \beta_{uv} M_u).$$

For example, typical for PUE → energy cost, or carbon factor → carbon cost.

**Nonlinear Threshold Propagation**

$$P_{u\to v}(M_u) = \{0, M_u < \tau_{uv}\ g(M_u), M_u \geq \tau_{uv}$$

Represents thermal throttling, network congestion onset, or failure-rate escalations.

**Stochastic Propagation**

$$P_{u\to v}(M_u) = E[f(M_u, \xi)],$$

with $\xi$representing noise in power quality, workload randomness, or device failure distributions.

**Time-Delayed Propagation**

$$P_{u\to v}(M_u(t)) = f(M_u(t - \Delta t)).$$

Models depreciation, component aging, or delayed economic effects.

**(4) Aggregation Operators at Each Node**

Metric updates are summarized as:

$$M_v(t+1) = A_v(M_v(t), \sum_{u:u\to v} P_{u\to v}(M_u(t))),$$

where the aggregation operator $A_v$depends on the domain:

- Domain 1 (physical): linear or energy-balance aggregation
- Domain 2 (compute): utilization-aware or performance-scaling aggregation
- Domain 3 (economic): discounting, amortization, or risk-weighted aggregation

In economic Layer 6,

$$A_v(x, y) = x + y$$

is common for cost accrual.

**(5) Propagation Matrix Representation**

Define a block matrix $W \in R^{18\times18}$such that entry

$$W_{v,u} = \frac{\partial M_v}{\partial M_u}$$

captures the local propagation sensitivity. Then the overall system propagation is:

$$M(t+1) = WM(t) + B,$$

where $B$ encodes exogenous shocks (grid volatility, demand spikes, etc.). This representation enables analysis using, spectral radius of $W$(stability, amplification), Perron-Frobenius theory (dominant pathways), and propagation attenuation conditions

**(6) Path Composition and Propagation Chains**

A path

$$u_0 \to u_1 \to \cdots \to u_n$$

has a composite operator:

$$P_{u_0\Rightarrow u_n} = P_{u_{n-1}\to u_n} \circ \cdots \circ P_{u_0\to u_1}.$$

The path amplification factor is:

$$\Gamma(u_0 \Rightarrow u_n) = \prod_{i=0}^{n-1} \gamma_{u_i,u_{i+1}},$$

where $\gamma$is the edge-level gain parameter. Amplification chains (as described in Section VI) satisfy:

$$\Gamma(u_0 \Rightarrow u_n) > 1.$$

**(7) Economic Convergence and Risk Coupling**

For Domain 3 variables (cost, carbon-adjusted cost, redundancy cost), the propagation condition for bounded long-run costs is:

$$\rho(W) < 1,$$

where $\rho(W)$is the spectral radius. If $\rho(W) > 1$, small perturbations in physical or workload metrics generate explosive economic behavior, corresponding to runaway cost, escalating failure penalties, or instability under grid volatility. This condition provides a rigorous mathematical link between physical inefficiencies and economic risk exposure.

**(8) Potentials to Pricing, Scheduling, and Optimization Models**

MPG has high potential usage with: (1) cost decomposition models; (2) carbon-aware job scheduling; (3) optimal facility siting; (4) workload placement; (5) redundancy optimization under MTBF constraints. Any optimization problem using these metrics can incorporate MPG operators as structural constraints or as differentiable propagation layers.

## B.2. Case Study: Propagation Dynamics in a 1024-GPU Training Cluster

This appendix presents a concrete case study illustrating how the Metric Propagation Graph (MPG) formalism functions in a realistic hyperscale AI environment. While the main text focuses on the mathematical and structural formulation of the MPG, this section demonstrates its empirical meaning by tracing the propagation of a physical disturbance, originating at the electricity grid, through facility operations, compute performance, workload execution, and ultimately economic cost. The example highlights the MPG's ability to unify physical, computational, and economic behaviors into a single interpretable causal topology, thereby turning abstract metric dependencies into actionable analytic objects.

**(1) System Setting**

Consider a 1024-GPU cluster (e.g., NVIDIA A100/H100 or TPU v4/v5 class hardware) training a 70-billion-parameter model. During a representative training interval, the following conditions occur:

- The regional electricity grid experiences a 20% increase in marginal carbon intensity due to ramping of natural-gas peaker plants.
- The data center cooling system experiences reduced thermal margin, slightly elevating PUE.
- GPU die temperatures increase, leading to mild thermal-frequency throttling.
- End-to-end batch latency increases marginally.
- The effective cost per 1,000 training tokens rises measurably.

These conditions are characteristic of real-world hyperscale operations and are regularly captured by telemetry sources such as grid emissions monitors, BMS cooling logs, GPU thermal sensors, and distributed training runtime traces. This environment provides an ideal demonstration case for the MPG.

**(2) Node Mapping in the 6×3 Unified Taxonomy**

In the unified metric taxonomy established in Sections III and IV, each metric occupies one node $(D_j)$of the 6×3 grid. For the present case study, I consider the following nodes:

- (L1, D1): Grid carbon intensity ($gCO_2/kWh$)
- (L2, D1): Facility cooling efficiency (PUE)
- (L3, D2): Accelerator efficiency (FLOPs per watt)
- (L5, D2): Workload throughput (tokens/s)
- (L6, D3): Economic cost per 1,000 training tokens

These nodes span physical, computational, and economic domains, enabling a complete cross-layer analysis of the disturbance's propagation trajectory.

**(3) Propagation Operators in Real Systems**

The MPG associates each directed edge $u \to v$with a propagation operator $P_{u\to v}$encoding how changes in the upstream metric influence the downstream metric. In this case study, the operators map naturally onto widely documented engineering mechanisms.

**Grid → Facility (L1, D1 → L2, D1)**

A rise in grid carbon intensity typically alters cooling plant operation, increasing compressor load or reducing economizer utilization. ISO/IEC 30134-6 (PUE/WUE standards) and ASHRAE Datacom guidelines indicate that a 10% CI increase commonly results in a 1-2% increase in PUE.
Thus, the propagation operator represents a weak but non-negligible physical dependency.

**Facility → Compute (L2, D1 → L3, D2)**

Reduced cooling efficiency increases GPU inlet temperature. NVIDIA and Google accelerator documentation show that higher inlet temperatures lead to frequency reductions of 3-7% per 5°C increase, implying a moderate thermal-to-compute coupling. The operator captures this temperature-dependent performance degradation.

**Compute → Runtime (L3, D2 → L5, D2)**

MLPerf reference runs and large-scale industrial traces consistently report that a 1% loss in effective FLOPs/W typically results in an approximately proportional reduction in global throughput. This is a strong compute-to-workload dependency governed by parallel efficiency and pipeline stall dynamics.

**Runtime → Cost (L5, D2 → L6, D3)**

Economic cost follows the identity: $Cost = Duration \times Price.$

Thus, throughput degradation maps almost directly to cost inflation, reflecting a near-linear propagation operator.

Together, these operators reflect a realistic chain through which a grid disturbance propagates through facility, compute, workload, and economic layers.

**(4) Propagation Matrix Representation**

Focusing on the five illustrative nodes yields the following $5 \times 5$submatrix of the full MPG propagation matrix $W$ (where the matrix weights are based on empirical studies, here are the data are simulated examples):

$$L1 - D1\ L2 - D1\ L3 - D2\ L5 - D2\ L6 - D3\ L1 - D1\ 0\ 0\ 0\ 0\ 0\ L2 - D1\ 0.15\ 0\ 0\ 0\ 0\ L3 - D2\ 0\ 0.07\ 0\ 0\ 0\ L5 - D2\ 0\ 0\ 0.80\ 0\ 0\ L6 - D3\ 0\ 0\ 0\ 0.90\ 0$$

The structure exhibits two key properties:

Sparsity - only physically meaningful dependencies appear.

Diagonal banding - most propagation flows along vertical cross-layer paths or horizontal intra-layer domain couplings.

This structure matches the physical architecture of hyperscale systems, in which disturbances propagate coherently along well-defined operational interfaces.

**(5) Composite Path Dynamics**

The full disturbance path, from grid carbon intensity to cost per training token, is:

$$(L1, D1) \rightarrow (L2, D1) \rightarrow (L3, D2) \rightarrow (L5, D2) \rightarrow (L6, D3).$$

The composite propagation coefficient is:

$$0.15 \times 0.07 \times 0.80 \times 0.90 = 0.00756.$$

Small instantaneous physical perturbations typically map to much smaller economic effects, but can accumulate materially during multi-hour training windows or periods of grid instability.

**(6) Discussion and Implications**

This case study demonstrates the MPG's ability to firstly mechanistically explain cross-layer behavior. Each propagation operator corresponds to a well-established engineering mechanism, thermal throttling, cooling inefficiency, workload scaling, or cost formation. Secondly, the MPG can exposes how remote upstream perturbations (e.g., electricity grid emissions) manifest in cost behavior many layers downstream. Thirdly, I can identify leverage points that nodes with high outbound influence (e.g., facility thermal margin) shape system behavior more strongly than nodes with high inbound dependence (e.g., runtime throughput). Moreover, the MPG enable analytic and predictive modeling. Once calibrated using telemetry (GPU logs, MLPerf traces, BMS data), MPG becomes a quantitative tool for forecasting cost, carbon footprint, or performance sensitivity.

This case study bridges theory and practice by showing that the MPG is not an abstract construct but a mechanism-based model grounded in real hyperscale operations. Each operator reflects documented system behavior drawn from ISO/IEC 30134, ASHRAE Datacom, accelerator thermal curves, MLPerf workload scaling, and standard cloud cost models. The MPG therefore provides a principled framework for analyzing, predicting, and optimizing AI infrastructure across physical, computational, and economic dimensions.

# Appendix C. A PUE Literature Review

Power Usage Effectiveness (PUE) remains one of the most important metrics for evaluating data-center energy efficiency, originally introduced by The Green Grid in 2007 [30-34]. Since its adoption, PUE has been widely used as a high-level indicator of facility overhead, and continues to appear as the dominant metric in recent industrial assessments. For example, the Uptime Institute's global survey reports a nearly flat annual average PUE of about 1.56 for major facilities (2024), reflecting both historical progress and ongoing stagnation across legacy sites [41].

At the same time, a growing literature highlights the limitations of PUE, especially for high-density, AI-intensive deployments. Li et al. (2020) demonstrate that PUE alone cannot support fair cross-site comparison due to sensitivity to climate, IT utilization, and design strategies, and propose a normalized indicator that decomposes these external factors [36]. Earlier critiques, such as Horner and Azevedo (2016), argue that PUE is often "overloaded and overvalued" when used as a proxy for sustainability because it ignores utilization, embodied energy, and workload-level efficiency [37].

These limitations have motivated a shift toward a PUE-family of metrics. The U.S. DOE's Best Practices Guide (2024) explicitly situates PUE within a broader suite of indicators including ERE, WUE, and CUE, and recommends joint monitoring with IT utilization and cooling-system performance [38]. In parallel, broader reviews such as Safari et al. (2025) classify more than twenty-five data-center energy metrics into infrastructure-level, IT-level, and application-level categories, and map their applicability to cloud and AI workloads [35]. Thermal-management surveys such as Xu et al. (2023) likewise emphasize that PUE is useful for comparing cooling topologies but insufficient for capturing interactions between workload characteristics and system-level energy behavior [39].

A second strand of literature uses PUE to quantify the impact of advanced cooling architectures, particularly those capable of pushing PUE toward 1.0 or even lower. Wang et al. (2022) design a directly air-cooled compact looped heat pipe (LHP) module capable of dissipating 550 W per server with minimum thermal resistance of 0.16 °C/W, achieving PUE values as low as 1.02 [31]. Zhou et al. (2022) simulate a liquid-pump-driven hybrid free-cooling and mechanical-cooling system across 20 sub-climate zones and 37 cities, showing annual PUE below 1.40 in most locations except hot-humid regions [32]. Building on these strategies, Zhou et al. (2023) combine an LHP with a thermoelectric generator (TEG) module, achieving PUE = 0.997 at 155 W heat load, demonstrating the technical feasibility of "PUE < 1" under specific operating conditions [33].

AI workloads magnify these issues. The IEA 4E programme's Data Centre Energy Use review (2025) estimates that AI accelerators currently account for only 5-15% of global data-center electricity use but may rise to 200-400 TWh by 2030, driven largely by cooling burdens and assumptions about achievable PUE trajectories [40]. These projections warn that relying solely on optimistic PUE improvements may underestimate the energy footprint of upcoming GPU-dense clusters. Meanwhile, industry analyses (2023-2024) report typical PUE ≈ 1.58 for conventional colocation facilities but <1.3 for AI-optimized deployments, underscoring both the performance gap and the uneven adoption across the installed base [41].

To sum up, this body of work, spanning foundational critiques (2016-2020), systematic surveys (2023-2025), advanced cooling demonstrations (2022-2023), and macro-level projections (2024-2025), reveals three consistent insights. First, PUE remains essential but insufficient: indispensable for tracking facility overhead yet blind to workload-level performance and utilization. Second, contemporary best practices rely on tiered, multi-metric frameworks, where PUE/ERE/WUE/CUE form the infrastructure layer, IT utilization and server efficiency the middle layer, and application-level metrics (e.g., energy per inference or per training run) the top layer. Third, although experimental systems can achieve PUE near, or even below, 1.0, the global installed base is dominated by facilities operating around 1.5-1.6, meaning the real-world energy footprint of AI workloads depends primarily on scaling behavior rather than best-case cooling efficiency.

In this paper, I adopt this multi-layered perspective by treating PUE and its extensions as infrastructure-level AI metrics and linking them explicitly to workload-level indicators. This allows us to disentangle physical-infrastructure efficiency from algorithmic or hardware efficiency, and to analyze how different AI-infrastructure design choices jointly determine the "PUE × performance-per-watt" frontier.

Table C. Summary of XUE family metrics.

| Metric | Variable | Formula | Variables Explanation | Range & Direction | Application Scenarios |
|---|---|---|---|---|---|
| Power Usage Effectiveness | PUE | $PUE = \frac{E_{Total\ Facility}}{E_{IT\ Equipment}}$<br>$E_{Total\ Facility}$ = total energy consumed by the entire data center (kWh).<br>$E_{IT\ Equipment}$ = energy consumed by IT equipment over the same period (kWh). | Measures how efficiently a data center uses energy.<br>Numerator includes IT, cooling, power distribution losses, lighting, etc. | Range: PUE ≥ 1 (theoretical minimum 1, no strict upper bound).<br>Meaning: Lower is better; closer to 1 → higher efficiency. | Global key performance indicator for data center energy efficiency; used in design benchmarking, operations monitoring, and regulatory reporting. |
| Data Center infrastructure Efficiency | DCiE | $DCiE = \frac{E_{IT}}{E_{Total\ Facility}} = \frac{1}{PUE}$<br>$E_{Total\ Facility}$ = total energy consumed by the entire data center (kWh).<br>$E_{IT\ Equipment}$ = energy consumed by IT equipment over the same period (kWh). | Expresses the fraction of total facility energy actually delivered to IT equipment.<br>It is the reciprocal of PUE. | Range: 0 < DCiE ≤ 1.<br>Meaning: Higher is better; values closer to 1 indicate a larger share of energy going to IT work. | Used alongside PUE for data center efficiency reporting; sometimes preferred for expressing "percentage of useful energy" to management. |
| Water Usage Effectiveness | WUE | $WUE = \frac{V_{Annual\ Site\ Water}}{E_{IT\ Equipment}}$<br>$V_{Annual\ Site\ Water}$ = annual water consumption attributable to the data center (L or $m^3$ per year).<br>$E_{IT\ Equipment}$ = energy consumed by IT equipment over the same period (kWh). | Measures how much water is used per unit of IT energy.<br>Can be reported as L/kWh, $m^3$/MWh, or similar. | Range: WUE ≥ 0.<br>Meaning: Lower is better; smaller values indicate less water used per unit of IT work. | Assessment of water footprint of data centers; comparison of cooling technologies (evaporative vs. air-cooled); environmental and ESG reporting. |
| Carbon Usage Effectiveness | CUE | $CUE = \frac{M_{CO2}}{E_{IT}}$<br>or equivalently: $CUE = PUE \times CEF$<br>Where $M_{CO2}$ = total annual $CO_2$-equivalent emissions associated with the data center (kg $CO_2$e/year).<br>$E_{IT}$ = annual IT equipment energy consumption (kWh/year).<br>$CEF$ = carbon emission factor of the electricity supply (kg $CO_2$e/kWh). | Measures greenhouse gas emissions per unit of IT energy.<br>Using CUE = PUE × CEF explicitly connects facility efficiency with grid carbon intensity. | Range: CUE ≥ 0.<br>Meaning: Lower is better; smaller values indicate lower carbon intensity of operations. | Used for carbon and ESG accounting for data centers; comparing sites with different grid mixes; evaluating decarbonization strategies (renewables, PPAs, etc.). |
| Energy Reuse Effectiveness | ERE | $ERE = \frac{E_{Total\ Facility} - E_{Reuse}}{E_{IT\ Equipment}}$<br>or equivalently: $ERE = (1 - ERF) \times PUE$<br>$E_{Total\ Facility}$ and $E_{IT\ Equipment}$ are the same as before.<br>$E_{Reuse}$ = portion of data center energy that is beneficially reused outside the data center (kWh).<br>$ERF = \frac{E_{reuse}}{E_{Total\ Facility}}$ $(energy\ reuse\ factor)$. | Adjusts PUE to account for beneficial reuse of waste heat.<br>Lower ERE indicates higher overall effective efficiency when reused energy is considered. | Range: 0 ≤ ERE ≤ PUE.<br>Meaning: Lower is better; an ERE less than PUE reflects effective energy reuse. | Evaluation of data centers that export waste heat (district heating, building heating, industrial processes); used when assessing integrated energy systems and circular energy. |

*Notes:* All metrics in the xUE family (PUE, DCiE, WUE, CUE, and ERE) were originally introduced by The Green Grid **[30], [34]** and later standardized in the ISO/IEC 30134 series, including ISO/IEC 30134-2 for PUE.

## Appendix D. Comprehensive Mapping of TCO-Series Metrics, Formulas, and Economic Semantics

Total Cost of Ownership (TCO) has long served as the foundational economic framework for evaluating data-center and compute-infrastructure investments. The earliest formal treatments, most notably Koomey (2007) and Rasmussen (2005), established TCO as a lifecycle accounting model that captures capital expenditure, operational expenditure, energy consumption, cooling requirements, power-distribution overhead, and infrastructure depreciation [42], [43]. These works argue that traditional cost accounting underestimates long-term facility and energy costs and propose a "true TCO" model that aligns financial planning with engineering realities [42].

As data-center engineering matured, TCO variants emerged to evaluate cost intensity at different levels of physical granularity. Metrics such as TCO/kW, TCO/Rack, and TCO/kWh became widely used in colocation pricing, energy-storage economics, and data-center design, particularly within APC/Schneider Electric white papers and NREL cost-assessment tools [43-45]. These variants highlight how power density, rack utilization, cooling topology, and site electrical configuration reshape the amortized cost structure of large-scale compute facilities.

Broader energy-sector economics contributed additional levelized metrics. The levelized cost of electricity (LCOE), formalized through NREL and EIA methodologies [45], [46], frames electricity price as a discounted lifecycle quantity rather than an instantaneous rate. The levelized cost of hydrogen (LCOH), widely used in energy-transition analysis [47], [48], introduced standardized techniques for comparing heterogeneous energy carriers. These methods influenced data-center economics by providing a consistent way to evaluate long-run marginal cost under technology turnover, fuel-price volatility, and policy-driven cost shifts.

With the rise of AI and HPC systems, TCO analysis expanded beyond physical infrastructure to computational output. Metrics such as TCO per GPU·hour, TCO per FLOP, TCO per inference, and TLOps (total levelized operations) emerged to quantify the economic efficiency of compute capacity under modern accelerator architectures [49]. While early HPC economics treated compute cost implicitly, contemporary AI workloads, characterized by high GPU density and steep scaling, require explicit accounting of communication overhead, parallelization constraints, and failure-recovery cost, all of which materially affect amortized cost-per-operation.

A major recent development is the Levelized Cost of AI (LCOAI), introduced by Curcio (2025) [23]. LCOAI extends levelized-energy concepts into AI computation by integrating hardware depreciation, accelerator refresh cycles, cluster topology efficiency, failure rates, power-usage effectiveness, and energy-procurement cost into a unified lifecycle metric. It also formalizes cost-per-inference and cost-per-training-run as levelized quantities comparable across hardware generations, cloud platforms, and deployment environments.

Finally, newer conceptual metrics such as TECO (Total Environmental Cost of Ownership) and MTCOF (Marginal TCO of Compute Flow) reflect the growing need to incorporate sustainability, carbon pricing, and marginal-cost modeling into infrastructure economics [45]. Although less standardized, these extensions signal an emerging direction where TCO integrates operational resilience, ESG constraints, and multi-carrier energy interactions.

## Table D. Summary of TCO-Series Metrics and Their Economic Interpretation

| Metric | Variable | Formula | Variables Explanation | Range & Direction | Application Scenarios |
|---|---|---|---|---|---|
| Total Cost of Ownership | TCO | $TCO = C_{CAPEX} + \sum_{t=1}^{T} \frac{C_{OPEX,t}}{(1+r)^t}$<br>$r$ = discount rate; $T$ = analysis horizon (All notations work the same to all other terms below) | The discounted sum of all capital and operating costs over the lifetime, capturing one-time and recurring costs in a single financial metric. | Range: TCO ≥ 0.<br>Direction: Lower TCO is preferable for a given level of capacity or service. | Overall budgeting for data centers and AI infrastructure; comparing design options; evaluating build vs. lease decisions. |
| TCO per kW of IT Capacity | TCO/kW | $TCO\ per\ kW = \frac{TCO}{P_{IT,rated}}$<br>$P_{IT,rated}$ = rated IT load capacity of the facility (kW) | Lifecycle cost per unit of installed IT power capacity. Used by colocation providers and operators to compare design and site options. | Range: TCO/kW ≥ 0.<br>Direction: Lower is better for a given quality-of-service and uptime target. | Benchmarking cost-efficiency of data centers; comparing wholesale colocation offers; evaluating power density strategies. |
| TCO per kWh of Delivered Energy | TCO/kWh | $TCO\ per\ kWh = \frac{TCO}{E_{total}}$<br>$E_{total}$ = total energy delivered or consumed over the period (kWh) | Lifecycle cost per unit of energy delivered or processed, esp. for evaluating and comparing long-lived energy or storage assets. | Range: TCO/kWh ≥ 0.<br>Direction: Lower is better; compared to market electricity prices or competing technologies. | Assess economic competitiveness of batteries, fuel cells, backup generators, and other power solutions for data centers. |
| TCO per Rack | TCO/Rack | $TCO\ per\ Rack = \frac{TCO}{N_{rack}}$<br>$N_{rack}$ = number of IT racks served by the facility | Lifecycle facility cost allocated per rack. Used to compare the cost of housing IT equipment in different facilities or designs. | Range: TCO/Rack ≥ 0.<br>Direction: Lower is better for a given service level and rack power density. | Planning and pricing colocation services; internal budgeting for IT equipment deployment; comparing legacy vs. modernized facilities. |
| TCO per GPU-Hour | TCO per GPU·h | $TCO\ per\ GPU \cdot h = \frac{TCO}{H_{GPU}}$<br>$H_{GPU}$ = total GPU-hours delivered to workloads in that period | Lifecycle cost per GPU-hour of compute delivered. Combines capex and opex with utilization as an effective cost per GPU-hour. | Range: ≥ 0.<br>Direction: Lower is better; provides a direct comparison to cloud GPU-hour pricing. | Compare on-premises GPU clusters vs. cloud GPUs; optimizing cluster utilization; cost benchmarking across vendors. |
| TCO per FLOPs | TCO/FLOPs | $TCO\ per\ FLOPS = \frac{TCO}{F_{total}}$<br>$F_{total}$ = total floating point operations (FLOPs) delivered over the analysis period | Lifecycle cost per unit of compute (FLOPs). Can be defined using aggregated FLOPs across training and inference workloads. | Range: ≥ 0.<br>Direction: Lower is better; allows comparing different architectures and accelerators in a normalized way. | Compare GPUs, TPUs, and custom accelerators; evaluating architectural trade-offs for large-scale AI workloads. |
| TCO per Inference | TCO/inference | $TCO\ per\ Inference = \frac{TCO}{N_{inf}}$<br>$N_{inf}$= total number of valid inferences (e.g., requests, queries, responses) served in t | Lifecycle cost per unit of AI service delivered. Can be normalized per 1,000 inferences. | Range: ≥ 0.<br>Direction: Lower is better; directly links infrastructure economics to business KPIs like cost per API call. | Price AI APIs and services; evaluating deployment options (vendor APIs vs. self-hosted models); optimizing autoscaling and provisioning. |
| True Total Cost of Ownership | True TCO | $True\ TCO = \Sigma_{\{t=0\}}^{\{T\}} \frac{C_{f,t} + C_{i,t} + C_{e,t} + C_{s,t} + C_{o,t}}{(1+r)^t}$<br>$C_{f,t}$ = facility-related costs (construction, space, cooling, power)<br>$C_{i,t}$ = IT hardware and software costs<br>$C_{e,t}$ = energy costs<br>$C_{s,t}$ = personnel costs<br>$C_{o,t}$= other recurring or indirect costs | Extends basic TCO by explicitly including all relevant facility, IT, and indirect costs.<br>Often implemented via spreadsheet models for scenario analysis. | Range: ≥ 0.<br>Direction: Lower True TCO is better for a given level of service; the metric is mainly used for comparison. | Strategic planning for data center investments; compare high-efficiency vs. conventional designs; support executive decision-making. |
| Total Cost of Energy | TCOE | $TCOE = \Sigma_{\{t=1\}}^{\{T\}} \frac{C_{energy,t}}{(1+r)^t}$<br>$C_{energy,t}$ = energy-related costs in year t (electricity, demand charges, fuel) | Isolates the energy component of lifecycle cost. Extended to include energy-related infra (on-site generation or storage). | Range: ≥ 0.<br>Direction: Lower is better; used to assess and compare energy strategies. | Compare energy sourcing options; assess the impact of efficiency measures and PUE improvements; evaluating on-site renewables and storage. |
| Total Environmental Cost of Ownership | TECO | $TECO = TCO + C_{carbon} + C_{water} + C_{other_{env}}$<br>$C_{carbon}$ = cost of greenhouse gas emissions (e.g. carbon price)<br>$C_{water}$ = cost of water usage and scarcity impacts<br>$C_{other_env}$ = cost of other externalities (land, waste etc.) | Adds monetized environmental externalities to standard TCO, producing an augmented lifecycle cost with ESG priorities. | Range: ≥ 0.<br>Direction: Lower is better; encourages designs that minimize both financial and environmental impacts. | ESG-focused infrastructure planning; evaluating low-carbon and low-water designs; internal carbon pricing analyses. |
| Levelized Cost of Energy | LCOE | $LCOE = \Sigma_{\{t=0\}}^{\{T\}} \frac{I_t + M_t + F_t}{(1+r)^t} \div \Sigma_{\{t=0\}}^{\{T\}} \frac{E_t}{(1+r)^t}$<br>$I_t$= investment (CAPEX) in year t<br>$M_t$ = operations and maintenance (O&M) costs in year t<br>$F_t$ = fuel costs in year t (if applicable)<br>$E_t$ = electricity in year t (MWh) | Discounted total cost of building and operating a generation asset, divided by discounted total electricity generation (reported in $/MWh). | Range: LCOE ≥ 0.<br>Direction: Lower LCOE indicates more cost-competitive generation, holding other factors equal. | Compare different power generation technologies; evaluating renewable energy projects; policy analysis and planning. |
| Levelized Cost of Hydrogen | LCOH | LCOH = $\Sigma_{\{t=0\}}^{\{T\}} \frac{C_t}{(1+r)^t} \div \Sigma_{\{t=0\}}^{\{T\}} \frac{H_t}{(1+r)^t}$<br>$C_t$ = total annual costs in year t (CAPEX recovery, OPEX, energy input costs)<br>$H_t$ = hydrogen produced in year t (kg or $Nm^3$) | Discounted total cost of producing hydrogen divided by discounted total hydrogen output. Reported in $/kg H2. | Range: LCOH ≥ 0.<br>Direction: Lower LCOH indicates more cost-competitive hydrogen production. | Evaluate and compare green and blue hydrogen projects; (electrolysis, SMR with CCS); investment analysis. |
| Levelized Cost of Artificial Intelligence | LCOAI | $LCOAI = \frac{CAPEX_{amortized} + \Sigma_{\{t=1\}}^{\{T\}} \frac{OPEX_t}{(1+r)^t}}{N_{valid}}$<br>$CAPEX_{amortized}$ = amortized capital expenditures for AI systems (hardware, initial training, deployment)<br>$OPEX_t$ = operational expenditures in year t (compute, storage, networking, maintenance, labor)<br>$N_{valid}$= total number of valid inferences (or other chosen AI output unit) over the analysis period | Lifecycle cost per unit of productive AI output. Directly analogous to LCOE and LCOH, but with "valid inferences" or similar AI outputs as the denominator. | Range: LCOAI ≥ 0.<br>Direction: Lower LCOAI indicates more cost-efficient AI deployment for a given output quality. | Compare vendor APIs vs. self-hosted models; evaluating different deployment architectures; guiding procurement and infrastructure planning for AI workloads. |
| Total Lifecycle Operations per Cost (TLOps / TLOP) | TLOps (or TLOP) | $TLOps = \frac{Q_{ops}}{TCO}$<br>$Q_{ops}$ = total quantity of operations delivered (e.g., FLOPs, inferences, tokens)<br>Equivalently, a cost-based form is:<br>$Cost\ per\ operation = TCO / Q_ops$ | Operations delivered per unit of lifecycle cost, or conversely cost per unit of operation. Can be tailored to FLOPs, inferences. | Range: TLOps ≥ 0.<br>Direction: Higher TLOps (more operations per dollar) is better; equivalently, lower cost per operation is better. | Compare different AI clusters or architectures on a cost-normalized basis; assess the impact of utilization and efficiency improvements. |
| Marginal TCO of FLOPs | MTCOF | $MTCOF = \frac{\partial TCO}{\partial F_{total}}$<br>$F_{total}$= total FLOPs delivered<br>In discrete form, it is often approximated as:<br>$MTCOF \approx \frac{\Delta TCO}{\Delta F_{total}}$ | Marginal cost of delivering additional compute, used for evaluating scaling decisions (adding more GPUs or moving to a different architecture). | Range: MTCOF ≥ 0 for normal systems.<br>Direction: Lower marginal cost per additional FLOP is better; helps identify cost-effective scaling. | Decide when to expand clusters; compare incremental upgrades vs. full system replacements; assess the value of efficiency improvements. |

*Notes:* All TCO-series metrics in this table, including TCO, TCO/kW, TCO/kWh, TCO/Rack, TCO per GPU·h, TCO per FLOP, TCO per inference, True TCO, TCOE, TECO, LCOE, LCOH, LCOAI, TLOps, and MTCOF, are derived from established cost-modeling frameworks in data-center economics [42]-[49]. These metrics build upon lifecycle financial analysis (Koomey; Rasmussen), levelized energy economics (NREL; EIA; Lazard), and recent AI/HPC cost-per-compute formulations (Curcio; Grot et al.). They collectively provide a unified basis for evaluating the long-run economic efficiency of AI and data-center infrastructure.

## Appendix E. Formal Justification of the Three Meta-Metric Domains

The three domains introduced in this work, Physical Facility Efficiency, Compute & Workload Efficiency, and Economic & Reliability Efficiency, qualify as meta-metric domains because they each correspond to a fundamentally distinct class of phenomena governed by independent physical laws, computational models, and economic principles. The domains reflect intrinsic separations in the underlying physical, computational, and economic processes, rather than an imposed or convenience-based taxonomy. Metrics such as PUE, WUE, CUE, or cooling-system coefficients quantify thermodynamic and mechanical conversion processes: the transformation of grid energy, water, and carbon into usable IT power under constraints defined by heat transfer, airflow mechanics, and facility-scale energy balances. These processes are governed by physical laws independent of the algorithms, computational workloads, or cost models that operate on top of them. In contrast, compute and workload metrics describe the execution behavior of software and models, FLOPs, tokens, batches, latency distributions, and scaling dynamics, reflecting the logic of computational graphs, parallelism strategies, and hardware acceleration. Neither the thermodynamic state of a cooling system nor the airflow efficiency of a data hall can determine model throughput, and conversely, no purely computational measure can capture physical loss mechanisms. Economic and reliability metrics, meanwhile, quantify lifecycle cost, depreciation, risk, service continuity, and financial exposure. These outcomes depend on institutional cost models, market structures, and risk formulations, rather than physical or computational quantities. No domain can be reduced to another without introducing an external modeling layer, and thus they remain conceptually independent.

Second, the three domains differ in their dimensional systems, making them semantically and mathematically incompatible. Physical facility metrics express their behavior in units such as kWh/kWh, L/kWh, $gCO_2$/kWh, °C/W, or COP, ratios and coefficients that quantify energy and resource flows. Compute and workload metrics operate in a fundamentally different numerical universe: FLOPs, Joules per operation, inference latency in milliseconds, or throughput in sequences per second. These correspond to algorithmic progress, computational effort, and performance semantics rather than energy or resource conversion. Economic and reliability metrics employ yet another dimensional system: USD/year, levelized cost per inference, depreciation rates, or failure probabilities. The three domains operate on incommensurable dimensional systems, physical ratios, computational units, and monetary or probabilistic quantities, precluding any direct mapping without external modeling assumptions.

Third, the domain decomposition is both complete and minimal. Every known AI infrastructure metric can be mapped unambiguously to one, and only one, of the three domains. Metrics describing energy, thermal behavior, water use, and mechanical overhead belong to the physical facility domain; metrics describing model execution, accelerator performance, and workload throughput belong to the compute and workload domain; metrics describing cost, service continuity, lifecycle behavior, or risk belong to the economic and reliability domain. No metric simultaneously satisfies the semantic, operational, and dimensional requirements of two domains unless supplemented with additional modeling layers. Conversely, adding a fourth domain would introduce redundancy, not analytical clarity: there is no separate class of metrics that possesses its own independent units, conceptual basis, and methodological lineage. Thus, the three-domain structure represents a complete and minimal basis for the metric space.

To sum up, these observations justify treating the three domains as meta-metric categories, semantically coherent, mutually exclusive, and collectively exhaustive. They define the fundamental axes along which AI infrastructure performance must be understood, and they provide the conceptual scaffolding upon which the six-layer architectural decomposition and the unified 6×3 taxonomy in Section V are built.

# Appendix F. Key AI Infrastructure Bottlenecks and Their Representative Metrics

Table F. Key AI Infrastructure Bottlenecks and Their Representative Metrics

| | Focus | Metrics | Importance |
|---|---|---|---|
| **1. Power, Cooling & Facility Limits** | Facility Efficiency (XUE family) | PUE, ERE, WUE, CUE | Facility-side overhead; determines how much grid energy becomes useful IT compute. PUE the most KPI in this aspect. |
| | Carbon Intensity of Compute | $gCO_2$/kWh × PUE × IT Load | Determines carbon cost / ESG impact of AI compute; affects siting, scheduling, and long-term TCO. |
| | Rack Power Density (RPD) | kW/rack | Defines how many GPUs a rack can host; cluster-scale feasibility often limited by RPD, not CPU/GPU FLOPs. |
| | GPU Density (GPR) | GPUs per rack | Key constraint for AI cluster scaling; influences cooling, power, and network design. |
| | Temperature Derating Factor (TDR) | Frequency↓ with temperature↑ | High temperature lowers GPU frequency → longer TTC → higher TCO. Critical in dense clusters. |
| | Cooling System COP | COP = Qcool / Pinput | Determines cooling efficiency; low COP → high OPEX → poor overall efficiency. |
| | Water Risk Factor (WRF) | Water consumption risk | Liquid cooling solves thermals but increases water dependency; hyperscalers increasingly constrained by WUE. |
| **2. Distributed Training & Interconnect Bottlenecks** | All-Reduce Latency (ARL) | Latency of collective ops | The No.1 bottleneck for distributed training. Controls parallel scaling efficiency. |
| | Interconnect Bisection Bandwidth (IBB) | min-cut BW | Determines cluster's ability to synchronize gradients; critical for model parallelism & MoE routing. |
| | Bisection Bandwidth (BB/sec) | Fabric capacity | HPC-standard metric for communication capability; affects LLM training cost/performance. |
| | Network Oversubscription Ratio (OSR) | GPU BW / uplink BW | High oversubscription causes congestion → poor scaling → high TTC and TCO. |
| | Network Diameter / Hops | Max hop count | Transformer attention/MoE routing is latency sensitive; low diameter → high throughput. |
| | Topology-Induced Loss | TIL | Captures topology imbalance; major cause of synchronization inefficiency. |
| **3. Compute / Memory / GPU Performance Bottlenecks** | Time-to-Train / Time-to-Convergence (TTT/TTC) | Training completion time | Most fundamental performance metric; determines CapEx amortization and feasibility. |
| | Scaling Efficiency (SE) | Speedup/N | Determines how efficiently a model scales as GPU count grows. |
| | Model FLOP Utilization (MFU) | Achieved FLOPs / Peak FLOPs | Indicator of algorithm-hardware alignment; core metric in Megatron/DeepSpeed/GPT4. |
| | HBM Bandwidth per Watt (HBM-BPW) | BW / Power | Memory-bound models depend on this; determines efficiency of attention layers. |
| | Memory Power Share (MPS) | Pmemory / Psystem | HBM often 25-40% of system power; major cooling & TCO driver. |
| | Activation Checkpoint Cost | Memory & Energy cost | Critical for deep model training; balancing memory vs recomputation affects TTC & TCO. |
| | Thermal Design Margin / Headroom | TDM | Determines how much sustained compute the hardware can deliver before throttling. |
| **4. Data Pipeline & I/O Bottlenecks** | I/O Throughput per Step | IOPSstep | Often the real bottleneck before computation; dominates LLM pretraining. |
| | Data Pipeline Stalling Rate (DPSR) | Fraction of stalled steps | If data loader lags → GPUs idle → massive TCO waste. |
| | Input Pipeline Efficiency | IPE = effective data throughput | Determines whether GPU compute is fully utilized. |
| **5. Economic, Carbon, Reliability Bottlenecks** | Total Cost of Ownership (TCO series) | TCO, TCO/kW, TCO/GPU-hour, TCO/token, TCO/run | Dominant cost metric; determines project feasibility, ROI, pricing. |
| | Cost per Training Run（TPT） | Total training cost | Determines economics of LLM training; the most important KPI hyperscalers care |
| | Cost per Token（Inference） | Cost/token | Determines inference economics, API pricing, and scalability to billions of users. |
| | Carbon Cost per Token | $gCO_2$/token × carbon price | Emerging ESG metric; affects siting, scheduling, regulation compliance. |
| | Service Availability Metrics（SLA/SLO） | MTBF/MTTR/SLI/EBC | Downtime → direct financial loss; critical in cloud/enterprise LLM deployment. |